\documentclass[a4paper,fleqn,usenatbib]{mnras}

\usepackage{newtxtext,newtxmath}

\usepackage{etoolbox}
\makeatletter
\patchcmd\@combinedblfloats{\box\@outputbox}{\unvbox\@outputbox}{}{%
   \errmessage{\noexpand\@combinedblfloats could not be patched}%
}%
\makeatother

\usepackage[T1]{fontenc}
\usepackage{ae,aecompl}

\usepackage{graphicx}	
\usepackage{amsmath}	
\usepackage{amssymb}	
\usepackage{euclid}
\usepackage{comment}
\usepackage{tablefootnote}
\usepackage[acronyms,nogroupskip,nopostdot]{glossaries}
\usepackage{glossary-mcols}
\usepackage{appendix}
\usepackage{xcolor}
\definecolor{mygreen}{RGB}{26,120,34}

\newglossary[cod]{codes}{cod}{ntn}{Codes}
\makeglossaries
\newglossaryentry{cod:CAMB}
{
  name=\texttt{CAMB},
  description={Code for anisotropies in the microwave background},
}
\newglossaryentry{cod:CLASS}
{
  name=\texttt{CLASS},
  description={Cosmological linear anisotropy solving system},
}

\newglossaryentry{cod:classy}
{
  name=\texttt{classy},
  description={Python wrapper for CLASS},
}

\newglossaryentry{cod:CosmicEmu}
{
  name=\texttt{CosmicEmu},
  description={Cosmic emulator based on the Mira-Titan cosmological simulation suite (successor of FrankenEmu based on the Coyote simulation suite).},
}
\newglossaryentry{cod:EuclidEmulator}
{
  name=\texttt{EuclidEmulator},
  description={Emulator code to emulate non-linear corrections (boost factors) to DM power spectra},
}

\newglossaryentry{cod:e2py}
{
  name=\texttt{e2py},
  description={Python wrapper for EuclidEmulator},
}

\newglossaryentry{cod:HACC}
{
  name=\texttt{HACC},
  description={Hardware/Hybrid Accelerated Cosmology Code},
}

\newglossaryentry{cod:HALOFIT}
{
  name=\texttt{HALOFIT},
  description={Analytical code to produce non-linear power spectra},
}

\newglossaryentry{cod:NGenHalofit}
{
  name=\texttt{NGenHalofit},
  description={Code to produce non-linear power spectra using a semi-analytical approach for large and a smoothing-spline-fit model for small scales},
}

\newglossaryentry{cod:PKDGRAV3}
{
  name=\texttt{PKDGRAV3},
  description={parallel k-D tree gravity code (version 3); Cosmological N-body tree code},
}
\newglossaryentry{cod:UQLab}
{
  name=\texttt{UQLab},
  description={Matlab-based uncertainty quantification framework},
}
\newacronym{BAO}{BAO}{Baryon accoustic oscillations}
\newacronym{CMB}{CMB}{cosmic microwave background}
\newacronym{DE}{DE}{Dark energy}
\newacronym{DM}{DM}{Dark matter}
\newacronym{ED}{ED}{experimental design}
\newacronym{EE}{EE}{EuclidEmulator}
\newacronym{EFHR}{EFHR}{Euclid Flagship High Resolution}

\newacronym{EOE}{EOE}{emulation-only error}
\newacronym{EoS}{EoS}{equation of state}
\newacronym{GR}{GR}{general theory of relativity}
\newacronym{LARS}{LARS}{Least angle regression-based selection}
\newacronym{LH}{LH}{Latin hypercube}
\newacronym{LHS}{LHS}{Latin hypercube sampling}
\newacronym{LV}{LV}{Large volume}
\newacronym{HOD}{HOD}{halo occupation distribution}
\newacronym{NLC}{NLC}{nonlinear correction}
\newacronym{PC}{PC}{principal component}
\newacronym{PCA}{PCA}{principal component analysis}
\newacronym{PCE}{PCE}{polynomial chaos expansion}
\newacronym{PCS}{PCS}{piece-wise cubic spline}
\newacronym{SPCE}{SPCE}{sparse polynomial chaos expansion}
\newacronym{THF}{THF}{Takahashi HALOFIT}
\newacronym{UQ}{UQ}{uncertainty quantification}
\newacronym{ZA}{ZA}{Zel'dovich approximation}

\title[The EuclidEmulator]{\textit{Euclid} preparation: II. The EuclidEmulator -- A tool to compute the cosmology dependence of the nonlinear matter power spectrum}

\author
[Euclid Collaboration]{Euclid Collaboration,
Mischa Knabenhans$^{1}$\;\href{https://orcid.org/0000-0002-2886-9838}
{\includegraphics[scale=0.75]{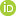}}\;\thanks{E-mail: mischak@physik.uzh.ch},
Joachim Stadel$^{1}$\,\href{https://orcid.org/0000-0001-7565-8622}{\includegraphics[scale=0.75]{orcid_16x16}}\,,
\newauthor
Stefano Marelli$^{2}$\,\href{https://orcid.org/0000-0002-9268-9014}
{\includegraphics[scale=0.75]{orcid_16x16}}\,,
Doug Potter$^{1}$\,\href{https://orcid.org/0000-0002-0757-5195}{\includegraphics[scale=0.75]{orcid_16x16}}\,,
Romain Teyssier$^{1}$\,\href{https://orcid.org/0000-0001-7689-0933}
{\includegraphics[scale=0.75]{orcid_16x16}}\,,
Laurent Legrand$^{3}$,
\newauthor
Aurel Schneider$^{4}$,
Bruno Sudret$^{2}$\,\href{https://orcid.org/0000-0002-9501-7395}
{\includegraphics[scale=0.75]{orcid_16x16}}\,,
Linda Blot$^{5,6}$\,\href{https://orcid.org/0000-0002-9622-7167}
{\includegraphics[scale=0.75]{orcid_16x16}},
Saeeda Awan$^{7}$,
\newauthor
Carlo Burigana$^{8,9,10}$,
Carla Sofia Carvalho$^{11}$,
Hannu Kurki-Suonio$^{12}$,
Gabriele Sirri$^{13}$
\\
$^{1}$Institute for Computational Science, University of Zurich, Winterthurerstrasse 190, 8057 Zurich, Switzerland\\
$^{2}$Chair of Risk, Safety and Uncertainty Quantification, Dept. of Civil Engineering, ETH Zurich, Stefano-Franscini-Platz 5, \\8093 Zurich, Switzerland\\
$^{3}$Kapteyn Astronomical Institute, University of Groningen, Landleven 12, 9747 Groningen, The Netherlands\\
$^{4}$Institute for Particle Physics and Astrophysics, Dept. of Physics, ETH Zurich, Wolfgang-Pauli-Strasse 27, \\8093 Zurich, Switzerland\\
$^{5}$Institute of Space Sciences (ICE, CSIC), Campus UAB, Carrer de Can Magrans, s/n, 08193 Barcelona, Spain\\
$^{6}$Institut d'Estudis Espacials de Catalunya (IEEC), 08034 Barcelona, Spain\\
$^{7}$Mullard Space Science Laboratory, University College London, Holmbury St. Mary, Dorking RH5 6NT\\
$^{8}$INAF, Istituto di Radioastronomia, Via Piero Gobetti 101, I-40129 Bologna, Italy\\
$^{9}$Dipartimento di Fisica e Scienze della Terra, Universit\`a di Ferrara, Via Giuseppe Saragat 1, I-44122 Ferrara, Italy\\
$^{10}$Istituto Nazionale di Fisica Nucleare, Sezione di Bologna, Via Irnerio 46, I-40126 Bologna, Italy\\
$^{11}$Instituto de Astrof\'isica e Ci\^encias do Espa\c{c}o, Faculdade de Ci\^encias, Universidade de Lisboa, Tapada da Ajuda, \\PT-1349-018 Lisboa, Portugal\\
$^{12}$Department of Physics and Helsinki Institute of Physics, Gustaf H\"allstr\"omin katu 2, 00014 University of Helsinki, Finland\\
$^{13}$INFN - Bologna, Istituto Nazionale di Fisica Nucleare, V.le Berti Pichat 6/2, 40127 Bologna, Italy\\}
\date{Accepted XXX. Received YYY; in original form ZZZ}

\pubyear{2018}

\newcommand{\vx}{\mathbf{x}}

\newcommand{\vX}{\mathbf{X}}

\newcommand{\cx}{\mathcal{X}}
\newcommand{\cy}{\mathcal{Y}}

\newcommand{\cm}{\mathcal{M}}
\newcommand{\ca}{\mathcal{A}}
\newcommand{\hompc}{\,h\,{\rm Mpc}^{-1}}
\newcommand{\mpcoh}{\,h^{-1}\,{\rm Mpc}}
\newcommand{\linv}{\ell^{-1}}

\newcommand{\valpha}{\mathbb{\alpha}}
\newcommand{\etaalpha}{\eta_{\valpha}}
\newcommand{\Psialpha}{\Psi_{\valpha}}

\newcommand{\corrOne}[1]{#1}
\newcommand{\corrTwo}[1]{#1}
\begin{document}
\label{firstpage}
\maketitle

\begin{abstract}
We present a new power spectrum emulator named \gls{cod:EuclidEmulator} that estimates the \corrOne{nonlinear correction} to the linear dark matter power spectrum \corrOne{depending on the six cosmological parameters $\omega_{\rm b}$, $\omega_{\rm m}$, $n_{\rm s}$, $h$, $w_0$ and $\sigma_8$}. It is constructed using the uncertainty quantification software \gls{cod:UQLab} using a spectral decomposition method called polynomial chaos expansion. All steps in its construction have been tested and optimized: the large high-resolution N-body simulations carried out with \gls{cod:PKDGRAV3} were validated using a simulation from the Euclid Flagship campaign and demonstrated to have converged up to wavenumbers $k\approx 5\hompc$ for redshifts $z\leq 5$. The emulator is \corrOne{based on 100 input cosmologies simulated in boxes of $(1250 {\rm Mpc}/h)^3$ using $2048^3$ particles}. We show that by creating mock emulators it is possible to successfully predict and optimize the performance of the final emulator prior to performing any N-body simulations. The absolute accuracy of the final nonlinear power spectrum is as good as one obtained with N-body simulations\corrOne{, conservatively}, $\sim 1\%$ for $k\lesssim 1\hompc$ and \corrTwo{$z\lesssim 1$}. This enables efficient forward modeling in the nonlinear regime allowing \corrOne{for estimation of cosmological parameters using Markov Chain Monte Carlo (MCMC) methods}. \gls{cod:EuclidEmulator} has been compared to \gls{cod:HALOFIT}, \gls{cod:CosmicEmu} \corrOne{and \gls{cod:NGenHalofit}}, and shown to be more accurate than these other approaches. This work paves a new way for optimal construction of future emulators that also consider other cosmological observables, use higher resolution input simulations and investigate higher dimensional cosmological parameter spaces.
\end{abstract}




\section{Introduction}
Next generation cosmological surveys of large-scale structure such as DES\footnote{www.darkenergysurvey.org} \citep{TheDarkEnergySurveyCollaboration2005TheSurvey}, Euclid\footnote{sci.esa.int/euclid} \citep{Laureijs2011EuclidReport}, LSST\footnote{www.lsst.org/lsst} \citep{LSSTScienceCollaboration2009LSST2.0} and WFIRST\footnote{wfirst.gsfc.nasa.gov} \citep{Green2012Wide-FieldReport} will exploit the highly nonlinear domain in order to vastly improve upon current precision estimates of cosmological parameters coming from \gls{CMB} experiments such as Planck \citep{Tauber2010PlanckMission,PlanckCollaboration:P.A.R.Ade2015PlanckParameters} and WMAP \citep{Bennett2003TheMission}. 
Euclid, to be launched by ESA in 2021, will measure the matter distribution in the Universe over most of its cosmic history \corrOne{(up to a redshift $z\approx 2.3$)}. Dark matter, dark energy and neutrino mass are currently the biggest challenges to modern physics. Euclid will be \corrOne{one of the first missions} to shed light on this dark sector, provided it manages to fully exploit the highly nonlinear scales of this large-scale structure. It is not just an observing challenge, but also a theory challenge that is laid down by these new large-scale structure surveys.

The theory delving into this highly nonlinear domain is extremely complex and computationally expensive as the desired level of accuracy is currently only achieved by cosmological N-body simulations.  
Such simulations are very expensive since both large simulation volumes and large numbers of particles are needed to reach the required precision. It is therefore mandatory to have theoretical tools able to much more rapidly predict cosmological observables on these small, highly nonlinear scales at an accuracy level of better than 1\% \citep{Huterer2005}. Even elaborate perturbation theory techniques break down below scales of \corrOne{$x\lesssim 10 \mpcoh$ or $k \gtrsim 0.6 \hompc$ \citep{Carrasco2014TheLoops}}. 
Fast, accurate and easy-to-use emulators like \gls{cod:EuclidEmulator} presented in this paper are critical to the success of large-scale structure surveys. 

Cosmic emulators provide a fast alternative to reliably predict cosmological observables, needing only a very small number of high precision N-body simulations during their construction. Recent examples include: {\em FrankenEmu}, based on the Coyote Universe simulations presented by Heitmann et al. in \citet{Heitmann2009, Heitmann2010, Lawrence2010,Heitmann2013}; \gls{cod:CosmicEmu} \citep{Lawrence2010CosmicEmu:Spectrum}, based on the Mira-Titan simulation suite discussed in   \citet{Heitmann2016,Lawrence2017}; and the {\em Aemulus} project introduced by DeRose, McClintock, Zhai et al. in \citet{DeRose2018TheCosmology,McClintock2018TheFunction, Zhai2018TheFunction}. Emulation makes use of pre-evaluated simulations for a relatively small set of cosmologies in a given parameter space. Having this data available, a surrogate model for a desired cosmological observable can be computed. This surrogate model computes the desired quantity for a given input cosmology within fractions of a second on a usual desktop machine.
Applications, such as Monte Carlo approaches for parameter space searches and forward modeling of cosmological observations, become feasible. This then also allows for likelihood sampling and thus for forecasting of Fisher matrices and Kullback-Leibler divergences \citep{Kullback1951OnSufficiency,Amendola2018}. Cosmological emulators can hence be used to accurately estimate the tightness of an error ellipsoid (referred to as ``Figure of Merit''), and thus are an important tool to maximize the science output of such large-scale projects.

Baryonic effects, such as cooling and feedback, complicate the study of matter clustering at medium and small scales because so far there is no self-consistent treatment of the relevant processes in the cosmological context. Recent hydrodynamical simulations report a suppression of power of the order of 10-30 percent at medium scales (${0.2<k<10 \hompc}$) followed by a strong enhancement at very small scales ($k>10 \hompc$) \citep{vanDaalen2011TheCosmology}, the latter is a consequence of baryon cooling and star formation in the halo centers. While most simulations reproduce this general trend, there is currently no agreement at the quantitative level. Some simulations predict a relatively small suppression affecting modes above ${k>1 \hompc}$ only \citep{Hellwing2016TheSimulation,Springel2018FirstClustering,Chisari2018TheSimulation}, others show a much stronger effect impacting modes above ${k>0.1 \hompc}$ \citep{vanDaalen2011TheCosmology,Vogelsberger2014PropertiesSimulation, Mummery2017TheStructure}.

The lack of agreement between different hydrodynamical simulations poses a serious challenge for future weak lensing and galaxy clustering surveys. Only if all baryonic effects can be controlled at the level of a few percent will it be possible to fully exploit the potential of future galaxy surveys like Euclid.
Recently, it has been shown that the amplitude of the baryon power suppression can be constrained with X-ray and Sunyaev-Zel'dovich observations of gas around galaxy groups and clusters \citep{Schneider2015ASpectrum,Mummery2017TheStructure,McCarthy2017TheCosmology}. This means that it is possible to come up with models to parametrize baryonic effects and calibrate them against observations \citep{Semboloni2011QuantifyingTomography,Zentner2013AccountingShear,Schneider2015ASpectrum,McCarthy2017TheCosmology}. These models can be encoded in a baryonic correction to the nonlinear power spectrum (sometimes referred to as the baryonic boost factor) that we hope to add to the analysis at later stage.

\corrOne{\citet{Davis1983ACorrelations,Kaiser1984OnClustersb,Bardeen1986TheFields}} and others have shown that galaxies cluster significantly differently than dark matter and hence a thorough understanding of this so called galaxy bias is crucial in order to compare observations to theoretical predictions based on DM simulations. While this bias is not part of the work presented in this publication, in the third paper of the Aemulus project series \citep{Zhai2018TheFunction} an emulation approach for the galaxy correlation function (and accordingly for the galaxy bias) is presented. \corrOne{They show that these quantities can be emulated by adding the relevant parameters to the cosmological parameter space, assuming that the \textbf{\gls{HOD}} approach is sufficient to model the galaxy bias.}

In this paper we present a new cosmic emulator for the nonlinear boost factor, i.e. the ratio between the nonlinear and the linear contribution of the matter power spectrum. This quantity is advantageous for three reasons: first, emulating the boost factor is more accurate than emulating the power spectrum directly. Recall that the linear power spectrum can be computed exactly using Boltzmann solvers like \gls{cod:CAMB} \citep{Lewis2000EfficientModels} or \gls{cod:CLASS} \citep{Blas2011} and hence the product of such a linear power spectrum and an emulated boost is more accurate than a directly emulated nonlinear power spectrum. Secondly, full transparency for all steps involved in the power spectrum estimation is maintained, as both the linear power spectrum and its \corrOne{nonlinear correction} are accessible in the entire emulation process. Thirdly, as the boost factor is multiplied by a linear power spectrum, the latter may feature additional physics that is not included in the nonlinear correction. As an example, a boost factor emulated based on the six parameter model (as laid out in this paper) still allows for a final nonlinear power spectrum that includes neutrino physics or general relativistic effects to linear order. Furthermore, the boost-factor approach provides a framework that can be easily extended at a later stage. For example, an additional boost describing the aforementioned baryon effects could be readily added to a future version of the emulator. For now we focus on the six parameter model inspired by Planck2015 \citep{Donzelli2015} including the baryon density $\omega_{\rm b}$, the matter density $\omega_{\rm m}$, the spectral index $n_{\rm s}$, the reduced Hubble parameter $h$, the \gls{DE} \gls{EoS} parameter $w_0$ and the variance $\sigma_8$ in a first step. We leave further parameters that quantify mostly deviations from standard $\Lambda$CDM models (as e.g. \corrOne{time varying} \gls{DE} \gls{EoS} $w_{\rm a}$, neutrino density $\omega_\nu$ or primordial non-Gaussianity of the local type $f_{\rm NL}$) to subsequent studies.

In contrast to prior emulators \citep{Heitmann2009,Heitmann2010,Lawrence2010,Heitmann2013,Heitmann2016,Lawrence2017,DeRose2018TheCosmology,McClintock2018TheFunction,Zhai2018TheFunction} that use {\em Kriging} \citep{Santner2013TheExperiments}, a Gaussian process interpolation technique, we use regression between the sample cosmologies using \gls{SPCE}, discussed e.g. in \citet{Blatman2011AdaptiveRegression}. \corrOne{Choosing this emulation technique we decrease the global maximal error of our emulator compared to a Kriging emulator}. As we will find in \autoref{Construction}, a sample of the cosmological parameter space (which in the field of uncertainty quantification, from now on abbreviated as UQ, is commonly referred to as the \textit{experimental design}) with 100 points being enough to achieve a global maximal emulation-only error (i.e. the relative error between the emulated boost spectrum and the boost spectrum computed from a full N-body simulation) below 1\%. In order to assess how the uncertainties on the input parameters affect the output observables, we use a state-of-the-art uncertainty quantification software called \gls{cod:UQLab} \citep{Marelli2014UQLab:Matlab}.

Further, we are the first to apply pairing and fixing techniques \citep{Angulo2016} together with an extension of the algorithm presented in \citet{Jing2005} on \gls{PCS} mass assignment \citep{Sefusatti2016} to pre-process the input cosmological simulations. This strategy allows us to drastically reduce numerical effects such as computational cosmic variance in the low $k$ regime and aliasing effects near the sampling Nyquist frequency.

In this work we mainly focus on the emulation strategy and how it can be optimized. As the power spectrum is a very fundamental quantity and because it is very natural to emulate, we choose it as our observable of interest. Emulation of other observables can and will be investigated in subsequent work.  

This paper is structured as follows: in \autoref{InputSims} the input simulations of the emulator and the applied optimization techniques are discussed. Then, in \autoref{Construction}, we investigate the actual construction and calibration of the emulator whose performance is assessed in \autoref{PerfAnalysis}. We summarize and conclude in \autoref{Conclusion}. We list the codes and acronyms used in this work together with short explanations in a glossary that can be found on page \pageref{ackns}.

\section{Input Cosmological Simulations}
\label{InputSims}
For the construction of an emulator, a full suite of high-quality cosmological simulations serves as the input data set. As will be discussed in \autoref{PerfAnalysis}, in our approach, the simulation errors are the dominant contribution to the uncertainties in the final emulated boost. As a consequence, the production of this data is not only very expensive but also challenging considering the tight bounds of 1\% on the power spectrum estimation set by the Euclid mission. Here, we describe a number of applied optimization techniques that allow us to reduce the computational time by roughly a factor of five compared to a standard N-body simulation approach without any decrease in the quality of the data.

\begin{figure}
  \includegraphics[width=\columnwidth]{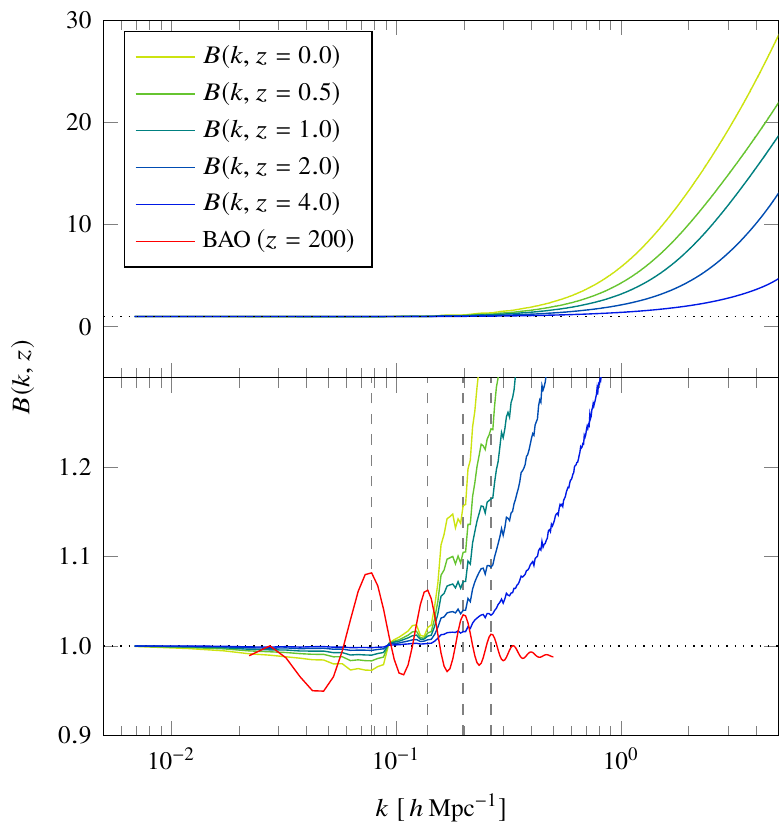}
  \caption{The \corrOne{nonlinear correction} to the power spectrum of the Euclid reference cosmology at different redshifts. While the overall behavior of the boost is smooth (top panel) a zoom in (bottom panel) reveals clear structure like the pre-virialization dip and the nonlinear suppression of the linear baryon acoustic oscillations (\gls{BAO}s; indicated by the red line in the bottom panel). In the bottom it is nicely visible how the nonlinear features become more and more distinct over time. The gray dashed vertical lines make clear that the minima of the nonlinear damping features coincide perfectly with the \gls{BAO} maxima.}
  \label{fig:Boost}
\end{figure}

The \gls{cod:EuclidEmulator} predicts the \corrOne{nonlinear correction} $B(k,z)$ of the dark matter power spectrum defined as
\begin{equation}
\label{eq:BoostDef}
B(k,z):=\frac{P_{\rm nl}(k,z)}{P_{\rm lin}(k,z)}\,,
\end{equation}
which divides the nonlinear by the linear dark matter power spectrum. An example \corrOne{nonlinear correction} is shown in \autoref{fig:Boost}, where the expected \citep{Eisenstein2005DetectionGalaxies,Crocce2008NonlinearOscillations} damping and broadening of the \gls{BAO} wiggles are evident. On $k\lesssim 0.1 \hompc$ there is a clear nonlinear suppression of power corresponding to pre-virialization \citep{Davis1977,Peebles1990,Jain1994}, which can also be understood as the nonlinear growth of voids at these scales.

The quality and performance of the emulator are highly dependent on the sampling of the cosmologies for which the N-body simulations are run. This sample of input cosmologies is called the \gls{ED}. In this section the simulation strategies for the computation of the experimental design are explained.

\subsection{Simulation of the experimental design}
\label{EDsimulaton}
The \gls{ED} was computed performing N-body simulations of the nonlinear matter power spectrum for a sample of 100 input cosmologies using the code \gls{cod:PKDGRAV3} \citep{Stadel2001,Potter2016PKDGRAV3:Code}. Each simulation started at the initial redshift $z_{\rm i}=200$ and evolved up to the final redshift $z_{\rm f}=0$ in 100 base time steps (smaller individual substeps are also used). Further details about the simulations will be discussed in \autoref{Simulations}. As our surrogate model emulates the nonlinear correction, the last step of the process in building the experimental design is to compute the \corrOne{nonlinear correction} for each simulated cosmology (further explanations in \autoref{Boost}). 

\begin{figure}
	\includegraphics[width=\columnwidth]{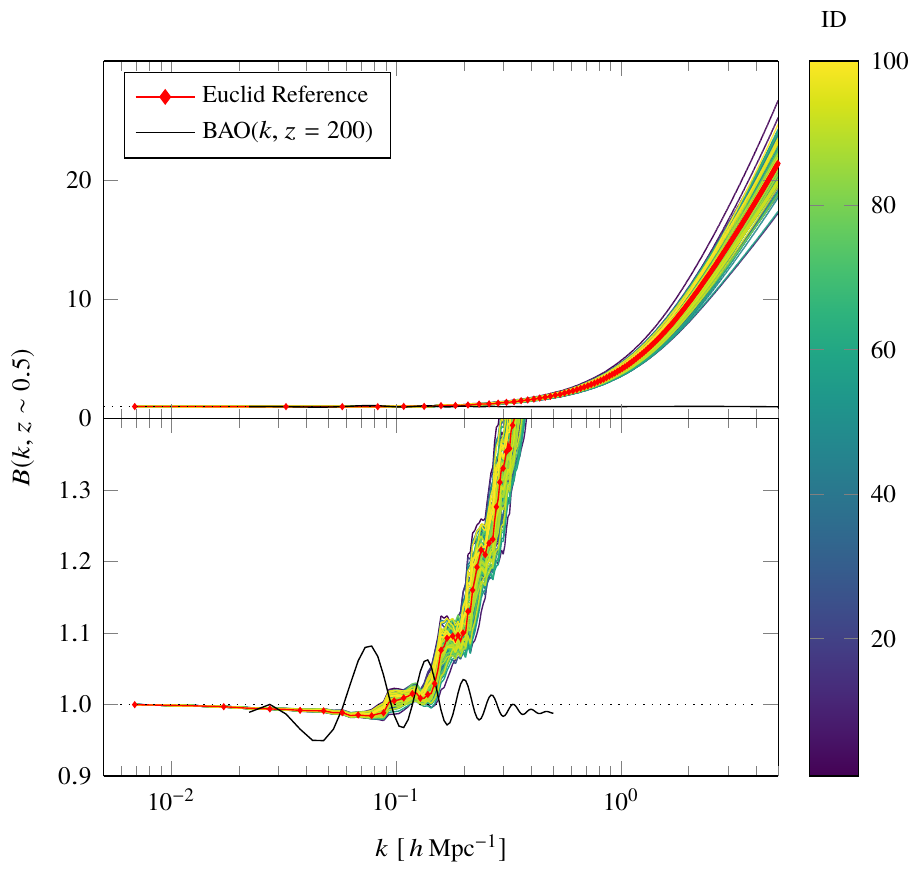}
	\caption{The experimental design used as input for the construction of the \gls{cod:EuclidEmulator}: \corrOne{nonlinear correction} curves for each input cosmology and the Euclid reference cosmology (red line with diamond markers) at output step 60 (roughly $z\sim 0.5$ depending slightly on the cosmology). The color bar labels the 100 different cosmologies in the \gls{ED} via their IDs. It is clearly visible that the Euclid reference cosmology is reasonably near the center of the parameter space as all \corrOne{nonlinear correction} of the input cosmologies scatter nicely around it. The black oscillatory curve in the lower panel indicates the \gls{BAO} at high redshift, as in \autoref{fig:Boost}.}
    \label{fig:ExpDesign}
\end{figure}

Convergence testing of the power spectrum (see appendix \ref{App:ConvTests}) was performed on the 
Euclid reference cosmology (\autoref{EucRefTable}) for which we had available a much higher resolution simulation (part of the Euclid Flagship simulation campaign, see \citealt{Potter2016}).
\begin{table}
\centering%
\caption{Cosmological parameters of the Euclid reference cosmology.}
\begin{tabular}{cccccc}
$\Omega_{\rm b}$ & $\Omega_{\rm m}$ & $n_{\rm s}$ & $h$ & $w_0$ & $\sigma_8$ \\
\hline
\hline
$0.049$ & $0.319$ & $0.96$ & $0.67$ & $-1.0$ & $0.83$\\
\hline
\label{EucRefTable}
\end{tabular}
\end{table}
However, the results at this particular reference cosmology are \textit{not} included in the ED. In \autoref{fig:ExpDesign} the set of 100+1 \corrOne{nonlinear correction} curves (including the \corrOne{nonlinear correction} of the Euclid reference cosmology) corresponding to the 60-th time step (equivalent to a redshift $z\sim0.5$) is shown.

\subsection{Cosmological N-body simulation}
\label{Simulations}
The matter power spectrum is well understood to linear order, i.e. in the regime where the dark matter overdensities in the Universe are small enough to allow a valid description in terms of linearized fluid equations. Apart from higher-order perturbative approaches there are no precise analytical means to calculate the nonlinear power spectrum. Several codes provide fast computations of these higher-order corrections \citep{Crocce2012,McEwen2016FAST-PT:Theory,Fang2017FAST-PTTheory} but all of them break down in the weakly nonlinear regime \citep{Carlson2009}. This is where N-body simulations come in. These codes are a direct first principles approach for \corrOne{simulating} the process of cosmic structure formation by numerically evolving the density field.

\subsubsection{The \gls{cod:PKDGRAV3} N-body code}

The main cost of producing an emulator of the nonlinear power spectrum is in performing the needed simulations over the cosmological parameter space. For this reason it is important to have available a fast and accurate N-body code. We have used the publicly available \citep{Potter2016PKDGRAV3:Code} N-body code \gls{cod:PKDGRAV3} by \citet{Potter2016, Stadel2001}. \gls{cod:PKDGRAV3} is a parallel fast multipole method (FMM) tree-code, which uses a block-step multiple time-stepping scheme for the integration of the equations of motion for the particles. It uses 5th order multipole expansions of the potential in calculating the force due to all the other particles, as well as for the calculation of periodic boundary conditions. 

\gls{cod:PKDGRAV3} has been validated \citep{Schneider2015} against two other well established N-body codes, namely \texttt{GADGET3} (for an older version of the code see \citealt{Springel2005TheGADGET-2}) and \texttt{RAMSES} \citep{Teyssier2010RAMSES:Code}. 
From this comparison we know that the absolute accuracy of power spectra generated with \gls{cod:PKDGRAV3} is better than $1\%$ for $k\lesssim 1\hompc$ and \corrTwo{$z\lesssim 1$} (at $z=0$, $1\%$-accuracy is achieved up to $6\hompc$). \gls{cod:PKDGRAV3} is very memory efficient, allowing for large simulations to fit on a relatively small number of nodes. In our case the $2048^3$ simulations fit comfortably on 16 nodes. Each simulation on 16 nodes (each node having 64 GB of RAM, 16 cores and no GPU) took almost exactly 3 days to complete. This adds up to 190\,000 node-hours to complete all $2\times100$ simulations used as input for our emulator (the factor of 2 comes from the pairing \& fixing described in \autoref{sec:FP}).

\subsubsection{Pairing \& fixing of initial conditions}
\label{sec:FP}
A common issue in power spectrum estimations in numerical simulations is the computational cosmic variance arising from the finiteness of the simulation box: very small Fourier modes (or equivalently very large physical distances) are undersampled. This leads to a loss of information or, put differently, to a large variance in statistical quantities like the power spectrum.
We do not want to include any contributions due to this effect within our emulator. 

We have used two techniques to reduce contributions from this sampling variance for the suite of input simulations. Firstly, it is possible to reduce these contributions at the linear level by computing the \corrOne{nonlinear correction} in a specific way, namely, by dividing the nonlinear power spectrum at redshift $z$ by the properly rescaled initial output power spectrum of the very same simulation. \corrOne{Here, a properly rescaled initial power spectrum is obtained by taking the linear power spectrum at redshift $z=0$ and scaling it back to the initial redshift $z_{\rm i}$ using the linear growth factor. However, due to mode-coupling in the non-linear evolution, the sampling variance still propagates somewhat from larger to smaller scales. This phenomenon adds a sample variance contribution to the power spectrum that remains despite the described division procedure}. Secondly, to further improve on this, we apply the method of  \corrOne{phase pairing and power spectrum amplitude fixing (hereafter ``pairing and fixing'')} described in \citet{Angulo2016}. This method is able to drastically reduce the computational cosmic variance and shall briefly be reviewed here: we use a fixed, linear input power spectrum $P_{i}$ (computed e.g. with \gls{cod:CLASS}) and draw the initial overdensity fields $\delta_{i, \rm lin}$ that can be decomposed into a magnitude $\vert\delta_{i, \rm lin}\vert$ and a phase $\theta_i$ according to the probability distribution function (pdf) given by
\begin{equation}
\label{eq:PFpdf}
P(\vert\delta_{i,\rm lin}\vert, \theta_i) = \prod_i \frac{1}{2\pi}\delta_{\rm D}(\vert\delta_{i,\rm lin}\vert-\sqrt{P_i})\,,
\end{equation}
with $\delta_{\rm D}$ being the Dirac delta function and the index $i$ labels the Fourier modes. This pdf identifies uniquely the magnitude of $\delta_{i,\rm lin}$ (it is ``fixed'') while the phase is still uniformly random between 0 and $2\pi$ such that one obtains
\begin{equation}
\delta_{i,\rm lin} = \sqrt{P_i}e^{i\theta_i}\,.
\end{equation}
\corrOne{For a comparison between paired-and-fixed simulations against traditional Gaussian random initial condition-based simulations, we refer the reader to appendix \ref{App:PF}}. Following this algorithm, we generate two initial conditions per set of cosmological parameters, both having the same magnitudes $\vert\delta_{i,\rm lin}\vert=\sqrt{P_i}$ but the phases being shifted by $\pi$ with respect to each other, i.e., we draw the first phase $\theta_{i,1}$ randomly and set $\theta_{i,2}=\theta_{i,1}+\pi$ for the second initial condition. 
For the generation of each initial condition we use the transfer function at $z = 0$ (from \gls{cod:CLASS}) and scale it back to high redshift ($z_{\rm i}=200$). Particle displacements are then set using the \gls{ZA}. We then perform a simulation for each of the two initial conditions and measure the power spectra. 

\corrTwo{\gls{ZA} was chosen over 2LPT for computing the displacement field due to the fact that a version of \gls{cod:PKDGRAV3} that correctly accounted for relativistic
fluids with 2LPT was not available at the time. Current 
developement versions of \gls{cod:PKDGRAV3} address this, and will allow to avoid the very high redshift starts using \gls{ZA} thereby minimizing systematic effects due to discreteness
without loss of accuracy. In principle 2LPT starts at lower 
redshift are favoured and will be considered in future work.} 
 
The resulting power spectra are then averaged (``paired'') and the \corrOne{nonlinear correction} is subsequently computed from the paired power spectra (for a deeper discussion of this \corrOne{nonlinear correction} computation, see \autoref{Boost}). We find that\corrOne{, on large and intermediate scales where computational cosmic variance poses a problem}, a \corrOne{nonlinear correction} computed with this algorithm is comparable to a \corrOne{nonlinear correction} coming from a power spectrum ensemble averaged over ten realizations. 
For a more detailed analysis of this algorithm and its performance we refer to \citet{Angulo2016,Pontzen2016InvertedVoids}. We found that we could reduce the computational effort by at least a factor of five using this method of pairing fixed simulations over conventional ensemble averaging. 

\subsubsection{4th-order mass assignment}
As the code \gls{cod:PKDGRAV3} evolves particles in a tree, their mass needs to be assigned to a grid whenever the power spectrum is computed.  The mass assignment scheme has a non-negligible impact on the quality of the power spectrum, particularly on nonlinear scales. While 2nd- and 3rd-order (cloud-in-cell and triangular shaped cloud) mass assignment schemes are widely used in simulations, we use 4th-order \gls{PCS} mass assignment as in \citet{Sefusatti2016}.
Although the time required for the mass assignment with this technique is increased, the errors in the power spectrum are substantially reduced. 

\subsubsection{Post processing: computing the nonlinear correction}
\label{Boost}
The main advantage of emulating the \corrOne{nonlinear correction} over full power spectrum emulation is that $B(k,z)=1$ on linear scales for all redshifts. This allows one to multiply it by a linear power spectrum that includes more physics on these large scales than can be explained by the relatively limited six-parameter model used for the \corrOne{nonlinear correction} computation itself (a prominent example is given by the super-horizon damping of the matter power spectrum captured, e.g., by the Boltzmann codes \gls{cod:CAMB} and \gls{cod:CLASS}). 
An added benefit is that emulating the logarithm of the \corrOne{nonlinear correction} appears to be almost an order of magnitude more precise than emulating the raw power spectrum, as is shown in \autoref{fig:ErrorMap} in \autoref{Construction}.

Having access to both the linear power spectrum from Boltzmann solvers like \gls{cod:CAMB} or \gls{cod:CLASS} and the power spectra from N-body simulations at all time steps, there are two different possible ways to compute the nonlinear correction:
\begin{enumerate}
\item Take the nonlinear power spectrum simulated by the N-body code and divide it by the linear input power spectrum computed with a Boltzmann solver like \gls{cod:CAMB} or \gls{cod:CLASS},
\item Divide the nonlinear power spectrum at redshift $z$ by the properly rescaled quasi-linear power spectrum at the initial redshift $z_{\rm i}$ of the N-body code.
\end{enumerate}
We follow the second approach for two reasons: firstly, this is the only approach where the \corrOne{nonlinear correction} is actually equal to 1 (as visible in \autoref{fig:Boost} and \autoref{fig:ExpDesign}) for low $k$-values as required by the argument stated above. This would not be achieved if one divided by the linear power spectrum computed with a Boltzmann solver.
Secondly, as mentioned in \autoref{sec:FP}, the former division already cancels out a considerable amount of computational cosmic variance.

We show in \autoref{ErrorPrediction} that only 100 cosmologies need to be simulated to achieve a maximal error of less than 1\% over the $k$ range of interest ${0.01\hompc\leq k\leq 5\hompc}$. We run two simulations per \gls{ED} sampling point in a $1250\mpcoh$ box with $2048^3$ particles, each with fixed initial conditions starting at redshift $z_{\rm i}=200$ and evolving to the present day ($z_{\rm f}=0$). We produce nonlinear 1D power spectrum outputs for 100 timesteps (equidistantly spaced in time) along the way. In a next step we average the power spectra over each pair of simulations ($P_1$ and $P_2$) and subsequently compute the \corrOne{nonlinear correction} spectrum at a certain redshift $z$ by dividing the averaged nonlinear power spectrum at redshift $z$ by the averaged nonlinear power spectrum at initial redshift. We thus compute
\begin{equation}
B(k,z)=\frac{\frac{1}{2}[P_1(k,z)+P_2(k,z)]}{\frac{1}{2}[P_1(k,z_{i})+P_2(k,z_{i})]}\left(\frac{D_{\rm 1LPT}(z_{\rm i})}{D_{\rm 1LPT}(z)}\right)^2
\end{equation}
instead of averaging the \corrOne{nonlinear corrections} themselves (in this equation, $D_{\rm 1LPT}$ denotes the scale independent 1LPT growth factor). In a comparison of these two calculation strategies they turned out to agree almost perfectly (to within less than 0.1\% over all wavenumbers of interest). Now we have an experimental design of $n_{\rm ED}=100$ \corrOne{nonlinear correction} spectrum sets each with $n_z=100$ different \corrOne{nonlinear correction} spectra (one for each redshift output step in the simulations) measured at $n_k$ = 2\,000 different linearly spaced $k$-points.

\subsection{Convergence of simulations}
As will be discussed below, the main contribution to the overall emulation error is due to the underlying simulations. We have performed a convergence test using different box sizes with edge length $L$ between 480 and $1920 \mpcoh$, with different particle numbers $N^3$ ranging from $1024^3$ to $2048^3$ and with different grid resolutions (once, twice or four times as many grid points as particles per dimension). For reference, two simulations have been used: a large volume simulation with a $(4000\mpcoh)^3$-box with $4096^3$ particles for assessing the minimally required simulation volume and a high resolution run with $8000^3$ particles in a $(1920\mpcoh)^3$ box to find the minimal mass resolution. We found that simulations with $L^3=(1250\mpcoh)^3$ and $N^3=2048^3$ particles (corresponding to a mass resolution of roughly $2\times10^{10}\;h^{-1} M_\odot$ per particle) have converged to the level of accuracy required, if a power spectrum measurement grid with roughly double this resolution is used. Using these specifications, we find that the simulated \corrOne{nonlinear correction} spectra have converged up to $k_{\rm max}=5.48\hompc$ for all redshifts $z\leq 5$ (reducing the number of $k$-points to 1\,100). For further details about the convergence tests, please refer to appendix \ref{App:ConvTests}.

\section{Emulator Construction \& Configuration} 
\label{Construction}
The emulated data is supposed to approximate simulations as accurately as possible. Accuracy, however, comes at the expense of higher cost in the construction of the emulator, or can result in an increase of the time and resources needed in the use of the emulator. In this section we will highlight the important aspects that influence the performance and the efficiency of the emulator and discuss how the \gls{EOE} can be reduced while keeping the overall costs for the construction of the emulator manageable. We define the \gls{EOE} as follows:
\begin{equation}
{\rm EOE}_{\textbf c}(k,z) = \frac{B^{\rm emulated}_{\textbf c}(k,z)}{B^{\rm simulated}_{\textbf c}(k,z)}-1\,, 
\end{equation}
where $k$ is the wavenumber, $z$ the redshift and $\textbf c$ stands for a cosmology for which the \corrOne{nonlinear correction} is evaluated. The steps involved in the construction of \gls{cod:EuclidEmulator} are: 
\begin{enumerate}
\item Definition of the cosmological parameter space, in our case, a 6-dimensional box over which a uniform prior is assumed,
\item \gls{LHS} of the parameter space,
\item N-body simulation of all cosmologies in the \gls{LH} sample,
\item Computation of the \corrOne{nonlinear correction} spectra (this data set in its entirety is called the experimental design),
\item \gls{PCA} of the nonlinear corrections,
\item \gls{PCE} of each individual \gls{PCA} coordinate, neglecting polynomial terms based on the sparsity-of-effects principle (hence \gls{SPCE} using \gls{cod:UQLab}),
\item Recombination of the principal components in a single emulator (using \gls{cod:UQLab} or our own stripped down C-code).
\end{enumerate}
For actually using \gls{cod:EuclidEmulator} to produce nonlinear power spectra, one only needs to combine step (vii) with a linear power spectrum generated by the \gls{cod:CLASS} or \gls{cod:CAMB} Boltzmann codes.

Redshift is not an emulated parameter and the \gls{ED} data matrix ${\mathbf D}$ contains a specific set of 100 \corrOne{nonlinear correction} spectra at different, cosmology dependent, redshifts (one for each output step of the simulations). To allow the computation of the \corrOne{nonlinear correction} at any requested $z$-value, we linearly interpolate between two adjacent \corrOne{nonlinear correction} spectra which bracket this redshift. \corrOne{By doing so, we commit the biggest error at the maximal redshift (because the input simulations are distributed less densely in redshift space towards higher redshifts) and maximal $k$-mode (as the change in \corrOne{nonlinear correction} is larger per $z$-interval for larger $k$-values) allowed by the emulator. We have tested that this maximal error is $\sim 0.6\%$. For all smaller $k$-modes and redshift values the error due to linear interpolation is smaller. Higher order interpolation over the data matrix ${\mathbf D}$ would make such errors at high $z$ and $k$ negligible, but since this consideration lies outside of the emulation strategy, we do not further consider it here.}

\subsection{Experimental design (ED) sampling}
The performance of the emulator crucially depends on how the \gls{ED} is constructed \citep{Blatman2011AdaptiveRegression}. The construction of the experimental design involves steps (i) to (iii). In this subsection these three phases shall be explained in more detail.

\subsubsection{Definition of the parameter space}
\label{DeltaRegion}
Similar to \citet{Lawrence2010}, our emulator is built upon the six parameter model including the following cosmological parameters:
\begin{itemize}
\item baryonic matter density parameter in the Universe, $\omega_{\rm b} = \Omega_{\rm b}h^2$,
\item total matter density parameter in the Universe, $\omega_{\rm m}=\Omega_{\rm m}h^2$,
\item reduced Hubble parameter $h$,
\item spectral index $n_{\rm s}$,
\item equation of state parameter of dark energy $w_0$,
\item power spectrum normalization $\sigma_8$,
\end{itemize}
where we assume a flat geometry of the Universe throughout ($\Omega_k=1$) such that the dark energy density parameter $\Omega_{\rm DE}$ is uniquely defined by the relation
\begin{equation}
\Omega_{\rm m}+\Omega_{\rm rad}+\Omega_{\rm DE}=1.
\end{equation}
These parameters are a subset of the parameters of the base $\Lambda$CDM cosmology from Planck \citep{Donzelli2015}. A key goal of the Euclid mission is to further constrain the \gls{DE} \gls{EoS} \citep{Amendola2018}. For this reason $w_0$ has been added to the investigated parameter space. Further important physical processes relevant for power spectrum measurement are, amongst many others, the effect of neutrinos on dark matter clustering or the impact of a time-dependent \gls{DE} EoS. Corresponding parameters have been included in the Mira-Titan Universe based \gls{cod:CosmicEmu} \citep{Heitmann2016, Lawrence2017} and will be included in future versions of \gls{cod:EuclidEmulator}.

We base our parameter box ranges on the Planck2015 best fit values mentioned in Table 4 in \citet{Donzelli2015}. For the parameters $\omega_{\rm m},\;n_{\rm s},\;h$ and $\sigma_8$ we use Planck-only data. However, as the constraining power of Planck for $\omega_{\rm b}$ and $w_0$ is significantly improved by combining it with external data, we use the combined best fit values for bounding the ranges of these two parameters. The upper and lower bounds are defined by $\mu\pm\Delta$, where $\Delta$ corresponds to $6\sigma$ quoted in \citet{Donzelli2015} for all cosmological parameters but $w_0$ ($\Delta_{w_0}=3.5\sigma$). The parameter ``box'' thus is finally defined as follows:
\begin{equation}
\begin{split}
\omega_{\rm b}&\in[0.0215, 0.0235]\,,\\
\omega_{\rm m}&\in[0.1306,0.1546]\,,\\
n_{\rm s}&\in[0.9283,1.0027]\,,\\
h&\in[0.6155, 0.7307]\,,\\
w_0&\in[-1.30, -0.70]\,,\\
\sigma_8&\in[0.7591, 0.8707]\,.
\end{split}
\end{equation}
We assume massless neutrinos for all cosmologies and take their energy contribution into account as a component of $\omega_{\rm rad}$. The photonic contribution to it is related to the \gls{CMB} temperature which we set to $T_{\rm CMB}=2.7255$ K. Our final $\omega_{\rm rad}$, including both photons and massless neutrinos, is cosmology dependent (via $h$). Since \gls{cod:CLASS} calculates this internally, we use its value for each \gls{cod:PKDGRAV3} simulation.

\subsubsection{Sampling}
\label{EDsampling}
The parameter space constructed above then has to be sampled in such a way that on the one hand one ends up with an experimental design containing only a relatively small number of points (otherwise the computational cost to produce the corresponding simulations explodes) and on the other hand the emulator built on top of these simulations must return highly accurate results. For now we assume no preliminary knowledge about the behavior of the emulated observable depending on the point in the parameter space. It is hence standard to use Latin hypercube sampling (LHS) \citep{McKay1979, Tang1993OrthogonalHypercubes}, which provides a fairly uniform coverage of the parameter space. For further discussion of statistical sampling techniques and their properties see \citet{Heitmann2009}, section 2.1 and references therein.

In order to perform a Latin hypercube sampling one needs to define the number of sampling points in advance. As we describe in \autoref{ErrorPrediction}, 100 sampling points are enough in order to construct an emulator that reaches the required accuracy in the output quantities. This step is performed using the statistics and uncertainty quantification software \gls{cod:UQLab}\footnote{http://www.uqlab.com}\citep{Marelli2014UQLab:Matlab}. As the construction of such a sample is a random process and not unique at all, we add an optimization step by generating $10^5$ different samples and choosing the realization for which the minimal distance (in Euclidean metric) between the sampling points is maximized (a classical maximin criterion, see e.g. \citealt{Johnson1990MinimaxDesigns}). By doing so we, e.g., avoid the unlikely event of all sampling points being aligned along the diagonal of the parameter space.

\subsection{Principal component analysis (PCA)}
Simulation data are usually noisy and suffers from non-physical, spurious numerical signals. We want the emulated data to be free from these problems which can be achieved by de-noising the input simulation data of the \gls{ED} using \gls{PCA}. The entire experimental design \corrOne{nonlinear correction} spectrum data set $\mathbf D$ can be represented as a $n_{\rm ED}\times(n_z\cdot n_k)$ matrix, where $n_{\rm ED}$ is the number of sampling points in the ED, $n_z$ is the number of output steps per simulation and $n_k$ is the number of wavenumbers considered for the power spectrum measurement. It turns out that the overall \gls{EOE} (the $L_{\infty}$ norm over all $k$ and $z$) is drastically reduced (by roughly an order of magnitude) if we do not store the \corrOne{nonlinear correction} values themselves into the data matrix $\mathbf D$ but rather use the logarithm thereof. The data matrix $\mathbf D$ is next decomposed into its $n_{\rm ED}$ principal components
\begin{equation}
\label{PCA}
\mathbf{D}=\sum_{i=1}^{n_{\rm ED}} \lambda_i(\omega_{\rm b},\omega_{\rm m},n_{\rm s},h,w_0,\sigma_8) {\rm PC}_i(k,z)\,,
\end{equation}
where the $\lambda_i$'s are the coordinates in the eigenbasis of $\text{cov}({\mathbf D})$ given by its eigenmodes  ${\rm PC}_{i}$ (vectors of length $n_z\cdot n_k$).

\subsection{The surrogate model}
\label{subsec:surmodel}
The main goal of emulation is to produce data in a simultaneously fast and precise way for all possible inputs. Therefore we choose to use polynomial chaos expansion (\gls{PCE}; a spectral representation on an orthonormal polynomial basis, see \citealt{Blatman2011AdaptiveRegression,Ghanem2003StochasticApproach,Xiu2006TheEquations,Xiu2010NumericalApproach}) in contrast to Gaussian process modeling (aka Kriging) as done by Heitmann et al., Lawrence et al. and also by Zhai et al. This strategy minimizes the global error, but comes at the expense of not exactly retrieving the simulation data at the input cosmologies.

Since according to \autoref{PCA} all information about the cosmological parameters is stored in the coordinates
$\lambda_i$, one only needs to find a surrogate for them. We therefore create a polynomial chaos expansion of each component separately. In the case of \gls{cod:EuclidEmulator} the expansion reads,
\begin{equation}
\label{PCFin}
\lambda_i(\omega_{\rm b},\omega_{\rm m},n_{\rm s},h,w_0,\sigma_8)\approx\sum_{\mathbf{\alpha}\in\mathcal{A}}\eta_\mathbf{\alpha}\Psi_\mathbf{\alpha}(\mathbf{x})\,, 
\end{equation}
where ${\mathbf{\alpha}=(\alpha_1,\dots,\alpha_6)}$ denotes a multi-index, $\Psi_{\alpha}$ the polynomial basis element  and  $\eta_\mathbf{\alpha}$ the corresponding coefficient. Here $\mathbf{x}=(x_1,\dots,x_6)$ is the vector of the six cosmological parameters, each mapped to the $[-1,1]$ interval. In practice the sum is truncated, making $\mathcal{A}$ finite and the above equation an approximation. The multivariate basis functions can be expressed in terms of normalized Legendre polynomials like so:
\begin{equation}
\Psi_{\mathbf \alpha}(\mathbf x) = \phi_{\alpha_1}(x_1)\phi_{\alpha_2}(x_2)\dots\phi_{\alpha_6}(x_6)=\prod_{l=1}^6\sqrt{2\alpha_l+1}P_{\alpha_l}(x_l)\,.
\end{equation}

The \gls{PCE} coefficients $\eta_\mathbf{\alpha}$ are computed with the least-angle regression (LARS) algorithm \citep{Efron2004LEASTREGRESSION,Blatman2011AdaptiveRegression}. 
This algorithm considers a set of candidate multivariate basis functions $\mathcal{A}_{\rm cand}$ defined by criteria related to the maximal total degree of polynomials $p$, the maximum interaction $r$ (number of non-zero values in the $\alpha$ vector of size 6), and a sparsity-inducing $q$-norm as follows:
\begin{equation}
\mathcal{A}_{\rm cand} =\left\{
\alpha: \; \left( \sum_{i=1}^6 \alpha_i^q\right)^{1/q} \le p, \quad
\sum_{i=1}^6 1_{\alpha_i \ne 0} \le r
\right\}\,.
\end{equation}
The least-angle regression then determines an optimal sparse set of polynomials $\mathcal A \subset \mathcal{A}_{\rm cand}$ such that a built-in error estimator on the truncated series \autoref{PCFin} is minimized. For a deeper discussion of the \gls{LARS} algorithm we refer to appendix \ref{app:LARS} or to the literature mentioned above.
The final performance of the emulator is tightly related to how the series is truncated, which terms are taken into account and which ones are dropped. The applied truncation scheme is a hybrid of hyperbolic and so called maximum interaction truncation \citep{Marelli2017UQLabExpansion}. For more elaborate instructions about how to compute a \gls{PC}E we refer to the referenced literature. 

\subsection{Optimizing and projecting emulator performance}
\label{ErrorPrediction}
A proper configuration of the emulator is key for good performance. While a misconfigured emulator can introduce \gls{EOE}s much larger than the simulation uncertainties (basically defeating the purpose of the emulator), a carefully configured surrogate model is able to introduce \gls{EOE}s so small that they are negligible compared to simulation errors. Such an emulator is thus capable of producing effectively simulation-quality results (but at much lower cost).

\subsubsection{The configuration space}
As we have discussed in the previous sections, the emulator construction process depends on various degrees of freedom (for a deeper discussion we refer to appendix \ref{App:UQ}):
\begin{itemize}
\item the number of sampling points $n_{\rm ED}$ in the experimental design,
\item the truncation parameters $p,q$ and $r$ characterizing the multi-index set $\mathcal{A}$,
\item the accuracy parameter $a$ defined as the fraction of the total variance captured by the principal components taken into account with respect to the total variance of the data. This is directly related to the number of principal components $n_{\rm PCA}$ taken into account.\footnote{Notice that the accuracy parameter $a$ is more fundamental than $n_{\rm PCA}$ since the latter additionally depends on further quantities as e.g. the size of the data set.}
\end{itemize}
As briefly described in appendix \ref{App:UQ}, the maximal polynomial order $p$ can be found following an iterative approach. This is done automatically by \gls{cod:UQLab} for every principal component separately and hence does not form part of the subsequent analysis. The remaining four parameters, though, need to be tuned carefully in order to optimize the emulator's final performance (i.e. balance its accuracy against its efficiency). We investigated this 4D parameter space on a grid given by
\begin{equation}
\begin{split}
\label{Eq:ConfigSpaceGrid}
n_{\rm ED} &= 10,25,50,75,100,250\,,\\
a &= 1-10^{-\kappa}\qquad\text{with }\kappa=1,2,\dots,10\,,\\
q &= 0.1,0.2,\dots,1.0\,,\\
r &= 2,3,4,5,6\,.
\end{split}
\end{equation}
For each of these 3\,000 grid points an emulator was constructed and used to make predictions that in turn were tested against a comparison data set. The relative error between prediction and comparison data was recorded. The most precise of all these 3\,000 emulators has then been filtered out under the requirement that it also be computationally efficient, i.e., under 0.1s on a single CPU core (comparable to the best Boltzmann codes).

\subsubsection{Comparison data set}
Notice that since we are interested in studying the \gls{EOE}, it is mandatory that both the \gls{ED} and the comparison data set are computed the same way (as otherwise differences in the computation strategy could contaminate the \gls{EOE}). As will become clear, this investigation requires more data than could be produced with N-body simulations. This is why we map out the error committed by emulating the \corrOne{nonlinear correction} using \gls{cod:CLASS} (version 2.6.3) and Takahashi's extension of \gls{cod:HALOFIT} \citep{Smith2003,Takahashi2012RevisingSpectrum} (hereafter abbreviated as $\rm THF$) as an alternative surrogate technique. For each cosmological parameter we have chosen 100 values equidistantly spread over the respective range resulting in a 6-dimensional lattice with $10^{12}$ points. As a computation of the \corrOne{nonlinear correction} spectrum for 1 trillion cosmologies is not feasible even with the halo model, we restrict the further analysis to the 15 coordinate planes of the parameter space (i.e. the planes of pairs of cosmological parameters) each of which is sampled by $10^4$ points, and a random sample of $10^4$ cosmologies in the bulk of the entire 6-dimensional space. For the resulting $16\times10^4$ cosmologies we compute the \corrOne{nonlinear correction} curves for redshifts $z=10^{-16},1,2$ and $5$ in the interval ${10^{-5}\hompc\leq k\leq10\hompc}$.

\subsubsection{Comparisons and tests}
\begin{figure*}
	\includegraphics[width=\textwidth]{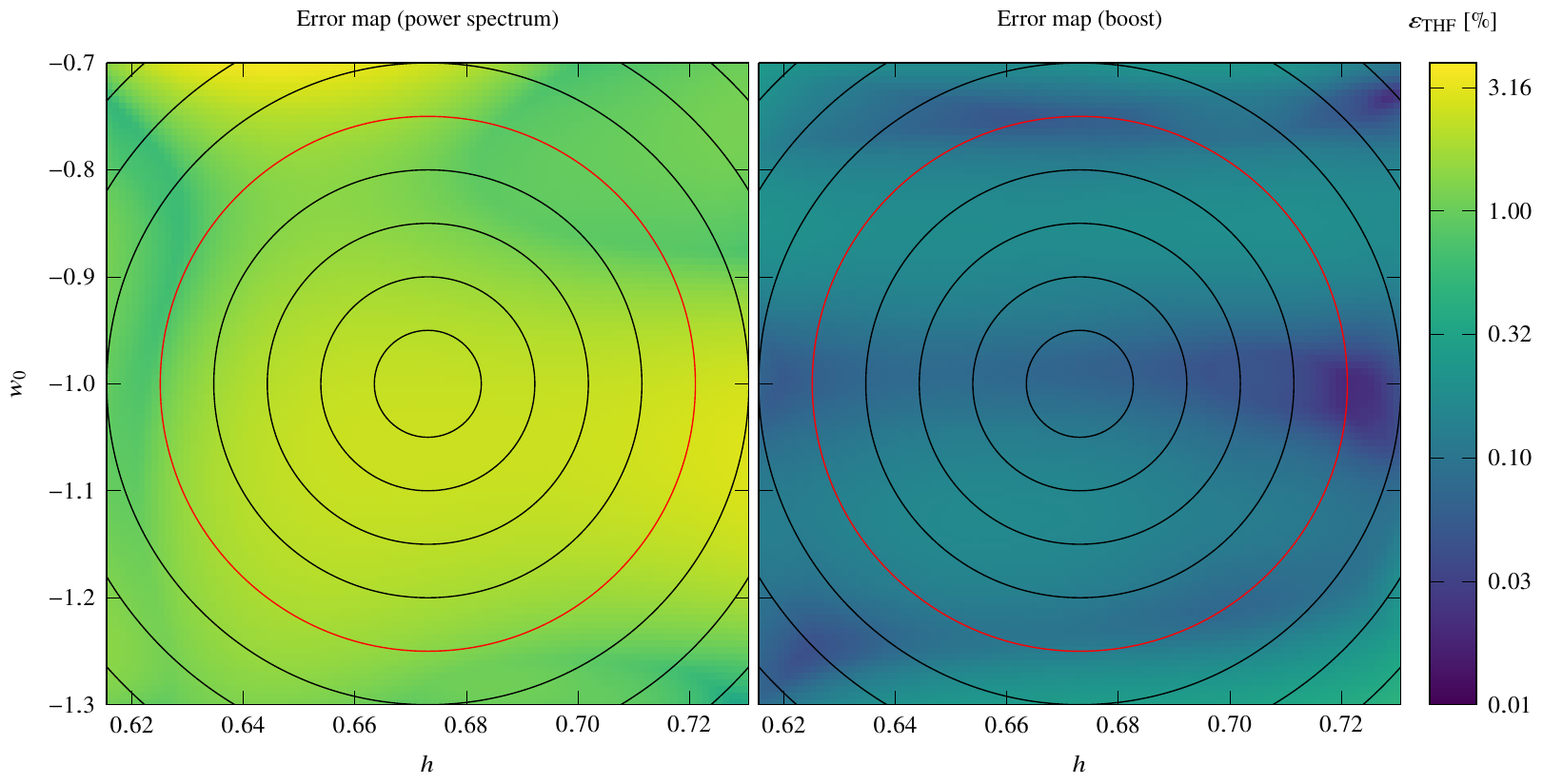}
	\caption{Error maps of the $(h,w_0)$-plane based on an \gls{ED} with $n_{\rm ED}=100$ for direct emulation of the power spectrum (left panel, similar to \citealt{Heitmann2010} and \citealt{Lawrence2017}) and the logarithm of the \corrOne{nonlinear correction} (right panel). We stress that logarithms are just used to construct the emulator and the errors shown here in both panels are based on comparisons of the full nonlinear power spectrum/nonlinear correction. For the \corrOne{nonlinear correction} emulation, the maximal error over all cosmologies is of order 0.5\% (in contrast to 3.7\% for direct power spectrum emulation) but if restricted to the $0.83\Delta$-ellipsoid (red ellipse), the maximal emulation-only error drops to below 0.2\% (direct power spectrum emulation: 2.75\%). The black circles indicate the different $\Delta$-ellipsoids (the innermost ellipse corresponds to the $0.17\Delta$, the second to the $0.33\Delta$ etc).}
    \label{fig:ErrorMap}
\end{figure*}

The EDs are as well computed using \gls{cod:CLASS} and Takahashi's halo model, one for each of the $n_{\rm ED}$-values, at the same four redshifts and for the same wavenumber intervals like in the comparison data set. We constructed emulators based on the logarithm of this \corrOne{nonlinear correction} data. To predict the \gls{EOE} (labeled as $\varepsilon_{\rm THF}$ to emphasize that it is based on the THF), we then evaluate the emulator at each lattice point of the cosmological coordinate planes, take the exponential of the emulation result in order to undo that logarithm and compare it to the data obtained directly from \gls{cod:CLASS}/THF. The relative error was subsequently computed and maximized over all redshifts and wavenumber values, and plotted as a density plot as shown in the example plot \autoref{fig:ErrorMap}. 
We thus get one such density plot for each coordinate plane. At this point we emphasize that while emulating $\log(B)$ instead of the \corrOne{nonlinear correction} itself is an essential technical step improving the final accuracy of the emulator dramatically, it has no effect on how the comparison is presented. For the comparison of directly computed and emulated \corrOne{nonlinear correction} spectra the logarithm has always been undone. 
Since the largest emulation errors are often found close to the boundaries of the parameter box, we exclude these by restricting the emulator to lie within the inner $0.83\Delta$-ellipsoid ($\Delta$ defined in \autoref{DeltaRegion}). The region outside of this is no longer considered in what follows.

\begin{figure}
	\includegraphics[width=\columnwidth]{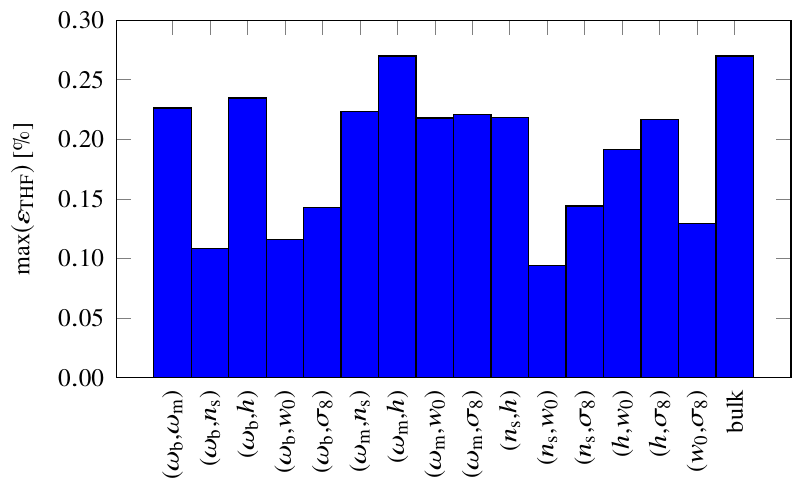}
	\caption{Maximal errors over all cosmologies in all coordinate planes and in the bulk of the space (restricted to the $0.83\Delta$ region). This bar chart shows that a comparison of error maps as in \autoref{fig:ErrorMap} looks similar for all coordinate planes: the maximal error of \gls{cod:EuclidEmulator} over all coordinate planes is about 0.27\% in contrast to roughly 1\% of the {\em FrankenEmu} \citep{Heitmann2010} and 4\% of the \gls{cod:CosmicEmu} \citep{Lawrence2017}.}
    \label{fig:5SigmaErrorBarPlot}
\end{figure}

\autoref{fig:5SigmaErrorBarPlot} shows the maximum error found within each of the 15 possible parameter planes, as well as over the entire parameter box (labeled {\em bulk}).
Clearly, all ${\rm max}(\varepsilon_{\rm THF})$ are comparable and considerably smaller than the maximal error reported in \citet{Heitmann2010} and \citet{Lawrence2017}.

\subsubsection{Cardinality of the experimental design}
\begin{figure}
	\includegraphics[width=\columnwidth]{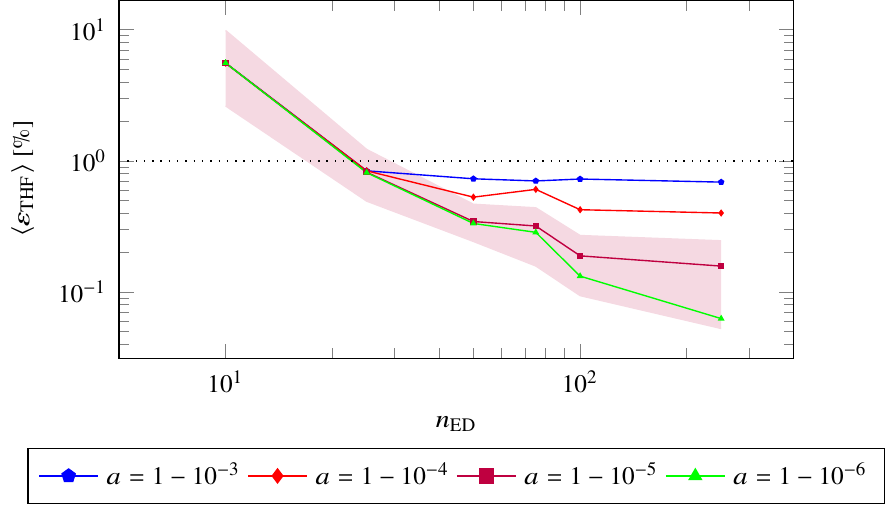}
	\caption{Dependence of the relative error on the size of the experimental design and the accuracy parameter $a$. The truncation parameters $r$ and $q$ have been set to 2 and 0.5, respectively, as these values are used for the construction of the actual \gls{cod:EuclidEmulator}. While the curves represent the mean of the maximal error (i.e. the error maximized over all $k$ and $z$ and then averaged over all cosmologies in the 15 coordinate planes and the bulk), the red shaded region additionally indicates the maximal and minimal maximal errors measured for $a=1-10^{-5}$. The curves plateau when the errors due to omission of terms dominates the errors coming from the sampling size.}
    \label{fig:NEDtest}
\end{figure}

We expect the maximal error to decrease as more cosmologies are included in the experimental design. This expectation has been tested over the range $10\leq n_{\rm ED} \leq 250$ keeping all the other parameters fixed. The result is shown in \autoref{fig:NEDtest}.
Notice that this plot suggests that as few as 50 cosmologies in the experimental design are enough to bring the maximal error within the $0.83\Delta$-ellipsoid down to below $1\%$ given the configuration used ($a=1-10^{-5}$). However, unlike the halo model, whose data is smooth, an N-body simulation will produce data with some noise. Techniques, such as \gls{PCA}, can reduce but not eliminate this noise contribution. Thus the overall maximal error of N-body simulation--based emulators can be larger than predicted by halo model--based emulators. For this reason we decided to build the \gls{cod:EuclidEmulator} on a conservative experimental design containing 100 cosmologies, reducing the expected relative error again by roughly a factor of 2.

It is important to notice that the relation between the parameter $a$ and the number of principal components taken into account is non-trivial and depends on $n_{\rm ED}$ as well as the nature of the data in the \gls{ED} itself (e.g. are the data noisy or not). As a result, the number of principal components is not constant along the curves in \autoref{fig:NEDtest}: the larger $n_{\rm ED}$, the more principal components are considered for a given parameter value for $a$.

\subsubsection{Truncation and accuracy parameters}
\label{Configuration}
The truncation parameters are $q~\in~[0,1]$ and $r~\in~\mathbb{N}$ (characterizing the multi-index set $\mathcal{A}$) together with the accuracy parameter $a\in[0,1]$ (or alternatively the number of principal components). We have constructed an emulator for each set $(a,q,r)$ in the grid defined in \autoref{Eq:ConfigSpaceGrid} using $N=100$ cosmologies in the experimental design and computed the maximal error as explained above. The goal now is to find the set $(a,q,r)$ that includes the least number of terms in the \gls{PC} expansion while keeping the \gls{EOE} low.

As this emulator shall be capable of evaluating many \corrOne{nonlinear correction} spectra within a second, we try to identify the most efficient emulator. We find that the accuracy parameter $a$ is the most dominant of those three parameters and that changing $q$ and $r$ does not have a significant effect as long as $r\geq 2$ and $q\geq 0.5$. We thus report the subsequent results always for $n_{\rm ED}=100$, $r=2$ and $q=0.5$ and only investigate the dependence on $a$ and $n_{\rm PCA}$, respectively.

We find that the smallest number of \gls{PC}s that have to be taken into account is $n_{\rm PCA}=4$ (corresponding to $a=1-10^{-3}$) as this leads to a maximal \gls{EOE} of just about 1\%. It is also possible to identify the emulator minimizing the maximal error which is achieved by setting $n_{\rm PCA}=26$ ($a=1-10^{-6}$) with $\max(\varepsilon_{\rm THF})\approx 0.1\%$ which is an order of magnitude smaller than the simulation uncertainty. Notice, though, that increasing $n_{\rm ED}$ further does not automatically improve the result: taking all principal components into account leads to an enhanced final error hinting at the fact that there is an optimal number of \gls{PC}s that can be taken into account.

Including more principal components will decrease the emulator performance as more terms have to be computed and the amount of input data for the emulator increases considerably. It is thus desirable to find a configuration that keeps the maximal \gls{EOE} well within the 1\% region but still leads to an efficient emulator. 

We have chosen to configure \gls{cod:EuclidEmulator} with the parameters
\begin{equation}
\begin{split}
n_{\rm ED} &= 100\,,\\
n_{\rm PCA} &= 11\qquad (a=1-10^{-5})\,,\\
q &= 0.5\,,\\
r &= 2\,.
\end{split}
\end{equation}
The conclusion of our Takahashi \gls{cod:HALOFIT} modeling is that our final \gls{cod:EuclidEmulator} should achieve a  maximal error of 0.27\% using only 100 different cosmologies, or 200 paired-and-fixed N-body simulations. 
We show in the next section that this appears to be confirmed for the final simulation-based emulator. 
Thus we can quickly and reliably predict the performance and minimize the computational cost of any future emulator, thereby maximizing the return of the entire N-body simulation campaign.

\section{Emulator Performance, Errors and Sensitivity to Parameters}
\label{PerfAnalysis}
\subsection{End-to-end tests of the emulation-only error (EOE)}
\label{E2E}
\begin{figure}
	\includegraphics[width=\columnwidth]{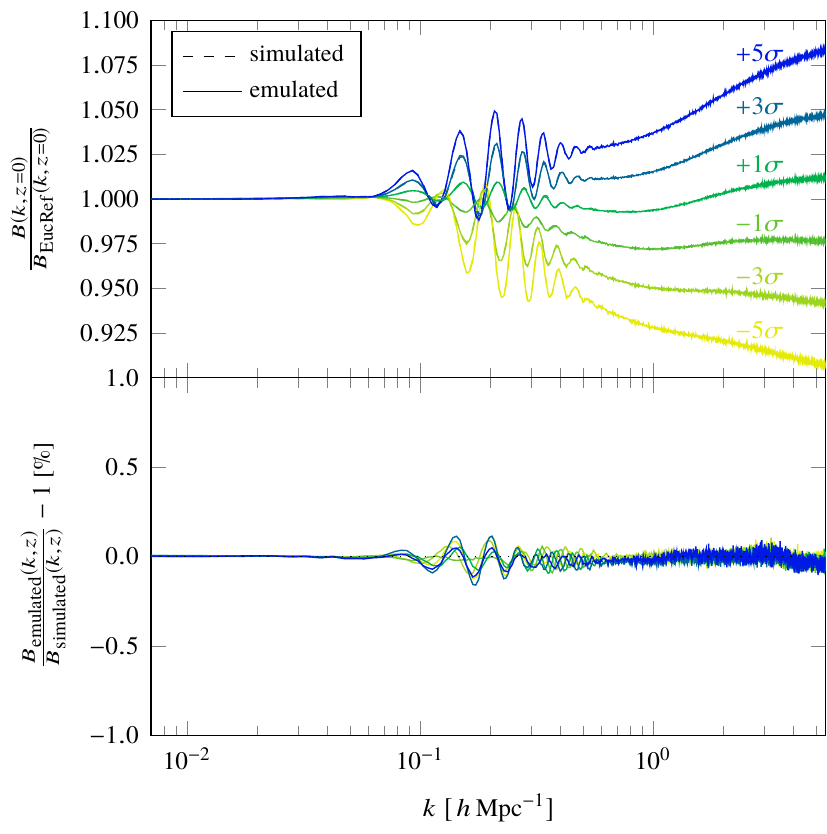}
	\caption{Comparison of the emulated (solid lines) and the paired-and-fixed (PF) initial condition-based simulated (dashed lines) \corrOne{nonlinear corrections} for six different cosmologies at $\pm1\sigma$, $\pm3\sigma$ and $\pm5\sigma$ away from the center of the $h$-parameter axis (see column 7 of \autoref{EOETable}). In the upper panel, corrections of the \corrOne{nonlinear corrections} themselves relative to Euclid reference (``EucRef'') cosmology are shown. No difference between the emulated and the simulated curves is visible. In the lower panel the relative differences between emulated and PF-simulated \corrOne{nonlinear corrections} are plotted. The entire $y$-range of the lower subplot corresponds to the accuracy tolerance regime of $\pm 1$\%, while the maximal error is roughly 6 times smaller than this upper limit. }
    \label{fig:hcompPlot}
\end{figure}

The predictions we have obtained in the previous section now have to be tested in an end-to-end manner for the actual \gls{cod:EuclidEmulator} based on real simulations. The test is performed along the 6 coordinate axes \corrOne{(varying only one parameter at the time)}. For each of them an N-body simulation of the cosmologies at $\pm1\sigma$, $\pm3\sigma$ and $\pm5\sigma$ (for $w_0$ we similarly chose $\pm0.58\sigma$, $\pm1.75\sigma$ and $\pm2.92\sigma$) from the center of the parameter range is run resulting in a test set of 36 reference simulations outside the experimental design used for the construction of the emulator. Then the emulator is executed at the very same cosmologies and compared to the simulations. In \autoref{EOETable} we report the relative errors maximized over the entire $k$-range, the redshift range and the $\sigma$-set (i.e. we report $\max[\text{EOE}(n\sigma),\text{EOE}(-n\sigma)]$):
\begin{table*}
\centering%
\caption{In this table we compare \corrOne{nonlinear correction} spectra computed with \gls{cod:EuclidEmulator} (\gls{EE}) on the one hand and with an emulator based on Takahashi's \gls{cod:HALOFIT} (THF) on the other hand to \corrOne{nonlinear correction} spectra of full N-body simulations for 36 cosmologies outside the experimental design over the entire $k$ and $z$ range. The numbers in the table correspond to the relative errors in percent between the emulated and simulated nonlinear corrections. We find that the errors predicted with the THF emulator are broadly consistent with the \gls{EE} errors. This implies that the configuration of the \gls{EE} based on a \gls{cod:HALOFIT} emulator is actually valid and that there is no need to make a huge investment to run N-body simulations just for the configuration study of an emulator. This result is particularly important for finding the size of the experimental design (see \autoref{fig:NEDtest}).}
\begin{tabular}{llllllllllllll}
\hline\hline
&\multicolumn{2}{c}{$\omega_{\rm b}$}&\multicolumn{2}{c}{$\omega_{\rm m}$}&\multicolumn{2}{c}{$n_{\rm s}$}&\multicolumn{2}{c}{$h$}&\multicolumn{2}{c}{$w_0$}&\multicolumn{2}{c}{$\sigma_8$}\\
\hline
&EE&THF&EE&THF&EE&THF&EE&THF&EE&THF&EE&THF\\
\hline
$\pm1\sigma$&0.081&0.081&0.093&0.091&0.082&0.076&0.078&0.119&0.082&0.076&0.086&0.097\\
$\pm3\sigma$&0.081&0.098&0.105&0.124&0.085&0.064&0.114&0.181&0.076&0.080&0.089&0.067\\
$\pm5\sigma$&0.073&0.062&0.166&0.220&0.121&0.069&0.145&0.127&0.081&0.062&0.089&0.096\\
\hline\hline
\end{tabular}
\label{EOETable}
\end{table*}
the overall maximal \gls{EOE} found is $0.145\%$ and thus much better than the error coming from the simulations and within the limit predicted by the \gls{cod:HALOFIT}-based error map. In \autoref{fig:hcompPlot} we explicitly compare simulated and emulated \corrOne{nonlinear correction} curves for six different cosmologies along the $h$-axis for redshift $z=0$ (plotted are the corrections relative to the Euclid reference cosmology). As will be established in the following \autoref{Sensitivity}, the Hubble parameter $h$ is one of the parameters the emulator is most sensitive to, even for higher-order principal components, and thus its variation should have a non-negligible effect. This is actually true as the six different cases are clearly distinguishable in the figure: While varying $h$ has almost no effect on linear scales, the curves corresponding to these six cases clearly deviate from one another on small scales\footnote{This makes it obvious, why it is important for surveys like Euclid to investigate the small scales. There is tremendous leverage on cosmological parameters in this regime.}. Yet, the emulated \corrOne{nonlinear corrections} coincide almost perfectly with the simulated ones \corrOne{that are based on paired-and-fixed initial conditions}. Though the relative differences (lower panel) do show a systematic around \gls{BAO} scale, these differences are negligible and the emulated data is effectively of simulation-quality. Note that the simulation of one of these \corrOne{nonlinear correction} curves takes about 2000 node hours while the corresponding emulated curve is computed within less than 50 milliseconds on a usual laptop. \gls{cod:EuclidEmulator} thus speeds up the data generation process by more than seven orders of magnitude compared to a classic N-body simulation with essentially no additional uncertainty due to emulation. 

\subsection{Sensitivity analysis}
\label{Sensitivity}
Sobol' indices \citep{Sobol1993SensitivityModels,Sobol2001GlobalEstimates} measure how sensitive the coordinates
$\lambda_i(\omega_{\rm b},\omega_{\rm m},n_{\rm s},h,w_0,\sigma_8)$ (introduced in \autoref{PCFin}) are to each single input parameter as well as to any of their interactions. For an introduction into Sobol' sensitivity analysis we refer to the previous references or to \citet{Marelli2017UQLabAnalysis,LeGratiet2016Metamodel-basedProcesses}.

\begin{figure}
  \includegraphics[width=\columnwidth]{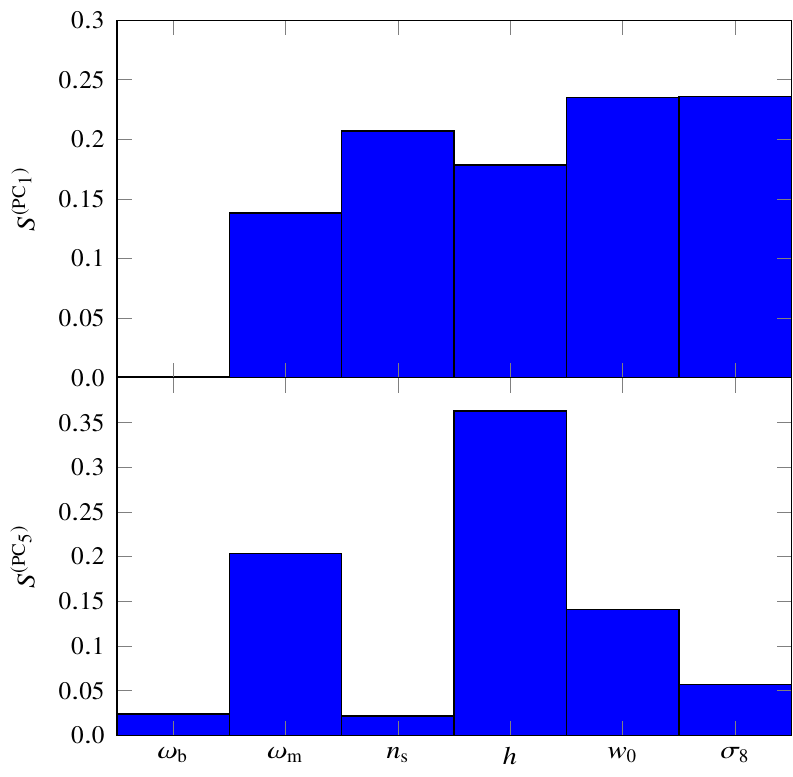}
  \caption{Sobol' index analysis plots. The bar plots show the individual first order Sobol' indices for each cosmological parameter for the first (upper panel) and the fifth principal component (lower panel). It is clearly visible that the baryon density parameter $\omega_{\rm b}$ has only a higher-order effect on the \corrOne{nonlinear correction} spectrum.}
  \label{SobolPlot}
\end{figure}

Sobol' indices are based on the Hoeffding-Sobol' decomposition, which states that any square-integrable function over a hypercube input parameter space can be cast as a sum of a constant, a set of univariate functions of each input parameter, another set of bivariate functions, etc. This decomposition is unique and the various terms are orthogonal with each other (with respect to the uniform probability measure over the hypercube). The variance of the output can then be apportioned to each input parameter, each pair, triplets, etc.: these contributions are called Sobol' sensitivity indices $S$. Although their classical estimation relies on costly Monte-Carlo simulations, Sobol' indices can be computed analytically from a polynomial chaos expansion as in \autoref{PCFin}, see \citet{Sudret2008GlobalExpansions}.
In our case, we get one such Sobol' expansion for each principal component. 
Each Sobol' index corresponds to the fraction of the total variance of the respective eigenvalue that is caused by the parameter(s) under consideration. The bigger this number, the more $\lambda_i$ depends on the considered set of input parameters. \autoref{SobolPlot} shows the first order (no interactions) Sobol' index plots for the first and the fifth principal component (see \autoref{fig:PCs} in appendix \ref{app:PCs}).
It is interesting to notice that the input parameter $\omega_{\rm b}$ has no leading order effect on the output nonlinear correction. The \corrOne{nonlinear correction} starts to show non-vanishing sensitivity to this parameter only at fifth order and higher. This is explained by the fact that baryons are treated as a background quantity that only come into the simulation data via the transfer function. As they are not directly evolved themselves in \gls{cod:PKDGRAV3}, their nonlinear contribution is only due to their mass which is taken into account in the $\omega_{\rm m}$ parameter.

\corrOne{
\subsection{Performance of \gls{cod:EuclidEmulator}}
\label{perf}
For the current implementation of \gls{cod:EuclidEmulator}, we have measured the execution times in three different setups (all times quoted were measured using one computing node):
\begin{enumerate}
\item the emulation of the full nonlinear power spectrum using the python wrapper of \gls{cod:EuclidEmulator} (called \gls{cod:e2py}) which in turn calls the python wrapper of \gls{cod:CLASS} (called \gls{cod:classy}) to compute the linear power spectrum. In this setup a wall-time of 0.37 seconds was measured.
\item the emulation of the nonlinear correction only using the C-code. For this task we measured a wall-time of 6 milliseconds. This time includes loading the information from the data table, calculation of cosmological quantities (e.g. the conversion from expansion factor to time), redshift interpolation and printing the results.
\item in a ``Monte Carlo setup'' (not yet available in the currently public version), i.e. the setup that would be used to actually perform an MCMC search of the parameter space. In this scenario one would load the data table only once and pre-compute the needed parts (dependent on the redshift) of the output data space ${\mathbf D}$ for the interpolation. This leaves calculating the \gls{PCE} and assembling the principal components for each MCMC step. In this case we measure an evaluation time of less than 5 microseconds (for $n_{\rm PCA}=11$). Notice that only in this setting we were able to measure the difference in wall-time between emulators taking different numbers of principal components into account as in the two previous cases this difference was unmeasurable compared to the total runtime. If we reduce $n_{\rm PCA}$ to 8, we measured 2.92 microseconds and 1.72 microseconds for $n_{\rm PCA}=2$.
\end{enumerate}
We stress that the current implementation of the code is not particularly optimized and any optimization at this point would be premature as clearly the biggest part of the calculation is spent in the computation of the linear power spectrum. This motivates the need for a comparably fast method to estimate the linear component, e.g. with a (separate) emulator. Clearly, this approach makes an MCMC search of the parameter space very efficient.
}

\subsection{Comparison to other fast prediction techniques}
We compare \gls{cod:EuclidEmulator} against two well-known alternative surrogate modeling tools: Takahashi's extension of \gls{cod:HALOFIT} and the \gls{cod:CosmicEmu} code based on the Mira-Titan Universe suite of simulations \citep{Lawrence2017} produced with the N-body code \gls{cod:HACC} described in \citet{Habib2016HACC:Architectures}. \corrOne{Moreover, we also compare \gls{cod:EuclidEmulator} against the very recent \gls{cod:NGenHalofit} \citep{Smith2018PrecisionUniverse}}. For these comparisons we use the Euclid Reference cosmology (i.e. the comparisons are out-of-sample tests) and they are performed on the level of power spectra, i.e. the \corrOne{nonlinear correction} curves computed by \gls{cod:EuclidEmulator} were multiplied with a linear power spectrum generated with the Boltzmann code \gls{cod:CLASS}. The result is then compared to the data from the other two predictors.
\subsubsection{Takahashi extension of \gls{cod:HALOFIT}}
\begin{figure}
  \includegraphics[width=\columnwidth]{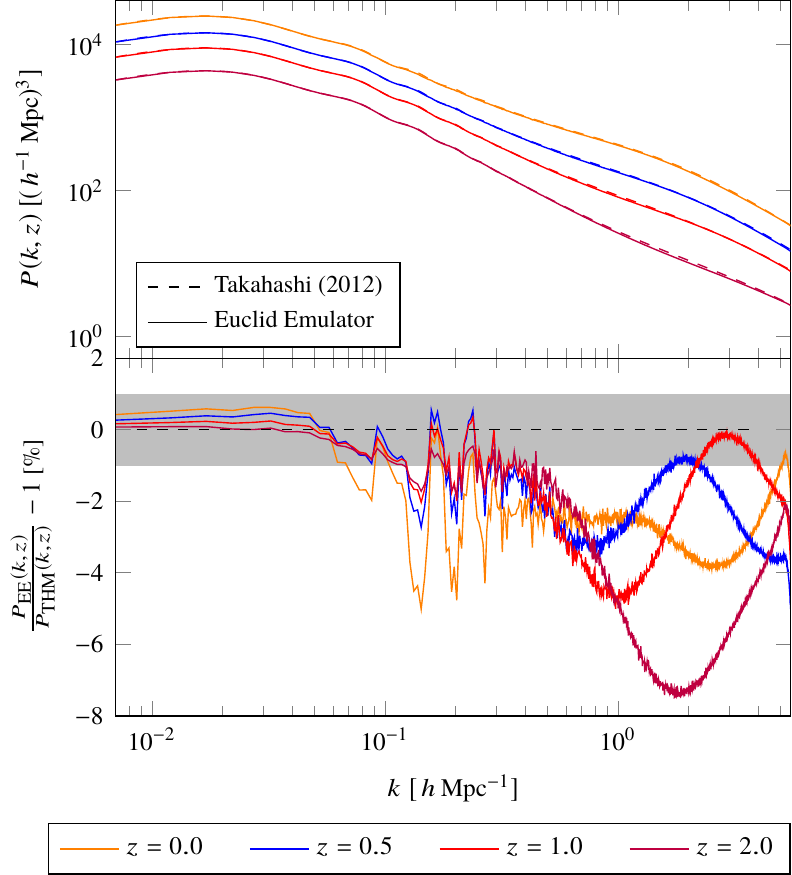}
  \caption{Comparison of nonlinear power spectra computed using \gls{cod:EuclidEmulator} (upper panel, solid curves) and the Takahashi model (upper panel, dashed curves). In the upper panel the absolute power spectra are shown while in the lower panel the relative error curves between the power spectra computed by the two different means are plotted. The agreement on large, linear scales is nearly perfect. On intermediate scales a distinct systematic coming from the \gls{BAO} signal is observed for all tested redshifts. This shows that the Takahashi model does not capture the \gls{BAO}s at the desired level of accuracy. The disagreement on the smallest scales is of order of several percent over all redshifts.}
  \label{EEvsTakahashiPlot}
\end{figure}

It becomes evident that \gls{cod:EuclidEmulator} is indeed able to correctly reproduce the linear regime of the power spectra (see \autoref{EEvsTakahashiPlot}). This is a big advantage of \corrOne{nonlinear correction} emulation over direct emulation of power spectra (as is clear from the comparison to the \gls{cod:CosmicEmu}, see \autoref{CompCosmoEmu}).
Further, one can see the distinct systematic wiggles at \gls{BAO} scales. They come from the fact that \gls{cod:HALOFIT} does not capture the nonlinear evolution of the \gls{BAO}s very well \citep{Heitmann2010}. On even smaller scales there is a clear disagreement between the Takahashi model and \gls{cod:EuclidEmulator} at the level of several percent. The differences, however, obey the uncertainty limits quoted in \citet{Takahashi2012RevisingSpectrum}. 
\subsubsection{\gls{cod:CosmicEmu} (Mira-Titan emulator)}
\label{comp2MT}
\begin{figure}
  \includegraphics[width=\columnwidth]{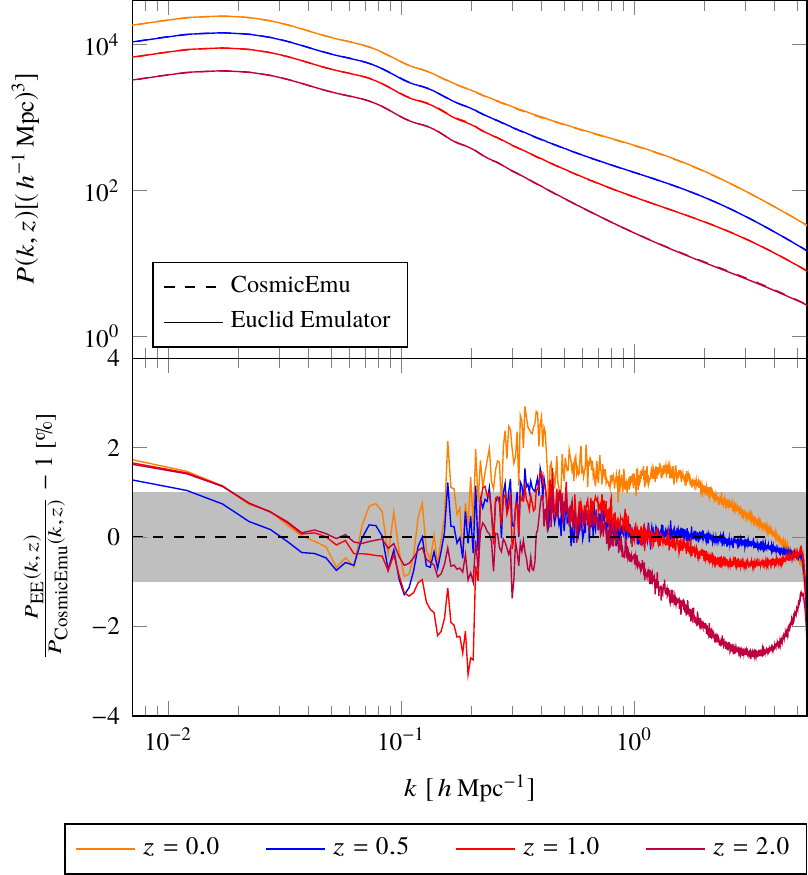}
  \caption{Comparison of nonlinear power spectra computed using \gls{cod:EuclidEmulator} (upper panel, solid curves) and the \gls{cod:CosmicEmu} code (upper panel, dashed curves). The relative errors between the power spectra computed with the two different approaches are again shown in the lower panel. On the largest scales there is a systematic disagreement originating in the \gls{cod:CosmicEmu} data (remember that on these scales \gls{cod:EuclidEmulator} is correct by construction). While on intermediate scales one observes disagreements of the order of a few percent as well, the agreement on small scales is very good over all redshifts.}
  \label{EEvsMTPlot}
\end{figure}

One observes a disagreement up to ${\sim}2\%$ between \gls{cod:CosmicEmu} and \gls{cod:EuclidEmulator} at linear scales (see \autoref{EEvsMTPlot}). This is partially explained by the fact that Lawrence et al. emulate the nonlinear power spectrum directly which introduces an error on all scales. As \gls{cod:EuclidEmulator} \corrOne{nonlinear correction} curves have to be multiplied with a linear power spectrum, the resulting nonlinear power spectrum matches linear theory on large scales by construction. As we use \gls{cod:CLASS} for the computation of the linear power spectrum, the resulting nonlinear curve does also contains the \gls{GR} corrections to the level given by \gls{cod:CLASS}. This is one of the biggest advantages of \gls{cod:EuclidEmulator}, but also comes at the expense of speed. Since direct emulation of the power spectrum circumvents the need for a Boltzmann solver, \gls{cod:CosmicEmu} is substantially faster: \corrOne{It takes \gls{cod:CosmicEmu} roughly 20 milliseconds to compute the nonlinear power spectrum}.

On intermediate and small scales the disagreement between \gls{cod:CosmicEmu} and \gls{cod:EuclidEmulator} is at most $3\%$ and thus consistent with the uncertainty bounds reported in \citet{Lawrence2017}.  
\label{CompCosmoEmu}
Summarizing, one can say that on large scales (${k<0.06\hompc}$) where the \gls{DM} clustering nicely follows linear theory, \gls{cod:EuclidEmulator} can be used to produce power spectra consistent with \gls{cod:HALOFIT} and Takahashi's extension well within the 1\% region. \gls{cod:CosmicEmu}, however, does deviate from the Takahashi model by a few percent on these scales; a consequence of the emulation strategy. On mildly nonlinear scales (${0.06<k<0.5\hompc}$) there is a certain disagreement between \gls{cod:EuclidEmulator} and both the Takahashi model and \gls{cod:CosmicEmu}, but of an entirely differing nature. While the deviation of Takahashi's model is systematic and correlated with the \gls{BAO} signal, the few percent differences between \gls{cod:EuclidEmulator} and \gls{cod:CosmicEmu} show an overall offset with redshift over these intermediate scales. On small scales (${k>0.5\hompc}$) \gls{cod:CosmicEmu} and \gls{cod:EuclidEmulator} are largely consistent in contrast to Takahashi's \gls{cod:HALOFIT} which systematically overestimates the nonlinear power by $4\%$ to $8\%$, depending on redshift. \corrOne{These observed discrepancies are broadly consistent with the ones shown in Fig. 5 of \citet{Schneider2015}. Only on the smallest scales (${k>3\hompc}$) and redshifts $z\geq 1$ there is also a mismatch in the comparison, which can be explained by the different mass resolution considered in that figure.}

\begin{figure}
  \includegraphics[width=\columnwidth]{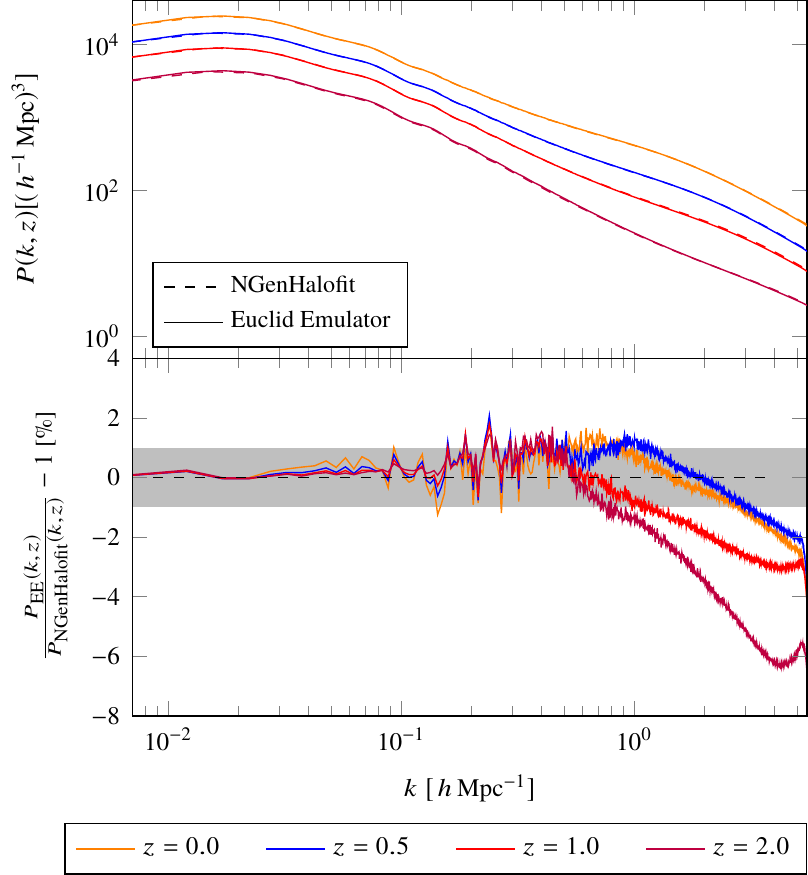}
  \caption{\corrOne{Comparison of nonlinear power spectra computed using \gls{cod:EuclidEmulator} (upper panel, solid curves) and the \gls{cod:NGenHalofit} code (upper panel, dashed curves). The agreement on linear and intermediate scales is very good with relative errors above $1\%$ at only a few $k$-modes around the BAO scale. However, there are substantial deviations on small scales which grow as one goes back in redshift.}}
  \label{EEvsNGenPlot}
\end{figure}

\corrOne{\subsubsection{\gls{cod:NGenHalofit}}
The agreement between \gls{cod:EuclidEmulator} and \gls{cod:NGenHalofit} is nearly perfect for large scales with $k\leq 0.1\hompc$ for all tested redshifts, as can be seen in \autoref{EEvsNGenPlot}. On intermediate scales the agreement is slightly above the $1\%$-level which is better than the corresponding results from the comparison to \gls{cod:CosmicEmu} or Takahashi's \gls{cod:HALOFIT}. On small scales, however, we observe a mismatch of up to $\sim 6\%$ (at $z=2$) which is also outside the bounds reported in \citet{Smith2018PrecisionUniverse}. \corrTwo{This disagreement may be explained by the fact that the D\"ammerung simulation suite  used to build \gls{cod:NGenHalofit} uses 2LPT initial conditions generated at redshift $z=49$ while the simulations used to construct \gls{cod:EuclidEmulator} are based on \gls{ZA} initial conditions (see discussion in \autoref{sec:FP}). We performed a comparison between \gls{ZA}-based simulations ($z_{\rm ZA}=200$) including radiation (the \gls{cod:EuclidEmulator} simulations) and 2LPT simulations ($z_{\rm 2LPT}=49$) without radiation (like the D\"ammerung simulations). At $z\sim2$ and $k\sim5\hompc$ we find an underestimation of power in the \gls{ZA} case compared to the 2LPT data at the level of roughly 3\%. While the two approaches agree perfectly on linear scales, the disagreement only becomes significant towards higher redshifts and higher $k$-modes (the agreement between \gls{ZA} and 2LPT is better than 1\% for all $z\lesssim 1$). The exact $k$-mode at which the maximal mismatch is located is resolution-dependent. This finding explains the excess mismatch we find in \autoref{EEvsNGenPlot}. This topic has also been discussed in \citet{Garrison2016ImprovingSimulations}, where the authors find that the 2LPT approach is the more accurate one.} The agreement is at the 3\%-level (or better) up to $z\sim1$ and out to $k\sim 5\hompc$.}

\label{comp2NGen}

\section{Conclusion}
\label{Conclusion}
Efficient and at the same time accurate estimation of nonlinear matter power spectra is crucial in order to exploit the full potential of cosmological surveys such as Euclid, DES, LSST and WFIRST. The Boltzmann solvers \gls{cod:CAMB} and \gls{cod:CLASS} are well-established as numerical tools to compute the linear matter power spectra. In this paper we hence focused on the \corrOne{nonlinear correction} in order to combine the strengths of the Boltzmann solvers and N-body simulation codes: the former include much more physics (such as \gls{GR} and baryonic physics) than any contemporary N-body code efficient enough to produce simulations of the size and resolution needed for current and upcoming surveys. The latter, however, are the only means by which structure growth on highly nonlinear scales can be studied. 
\gls{cod:EuclidEmulator}, presented in this paper, is a numerical tool that estimates the \corrOne{nonlinear correction} spectra of an input cosmology (respecting the predefined parameter boundaries of $0.83\Delta$ around the Planck2015 best-fit cosmology) at any redshift $z\leq5$ with an overall accuracy far better than 1\% based on only 100 pre-evaluated dark matter-only simulations performed with the N-body code \gls{cod:PKDGRAV3}. The emulation-only error is of order of a fraction of a percent and is hence dominated by the expected simulation errors (of order ${\sim}1$\% up to $k {\sim}1 \hompc$).

The accuracy of the emulation could be achieved by using well-tested statistical techniques from the field of uncertainty quantification: like \citet{Heitmann2010}, we have used a special sampling technique called Latin hypercube sampling (LHS) in order to guarantee that the resulting experimental design of input cosmologies covers the cosmological parameter space in a statistically uniform way. We then simulate the corresponding nonlinear responses with \gls{cod:PKDGRAV3}, using paired-and-fixed \citep{Angulo2016} initial conditions that drastically reduce computational cosmic variance in the simulations. In contrast to the Coyote universe, the Mira-Titan universe and the Aemulus project emulators by \citet{Heitmann2010,Lawrence2017,Zhai2018TheFunction} respectively, we employ a regression strategy called sparse polynomial chaos expansion in order to surrogate model the \corrOne{nonlinear correction} spectra. Our \corrOne{nonlinear correction} approach leads to very accurate emulation of the nonlinear matter power spectrum, but additionally requires the linear power spectrum, calculated from \gls{cod:CLASS} or \gls{cod:CAMB}. 

The emulator itself depends on a set of numerical parameters which need to be configured properly. To perform this configuration we predict the emulator performance for a given set of emulation parameters (such as the size of the experimental design, the number of principal components taken into account, the truncation of the polynomial chaos series) using \gls{cod:HALOFIT} input data. We \gls{LH} sample experimental designs of different sizes and compute mock emulators based on \gls{cod:HALOFIT}/\gls{cod:CLASS} \corrOne{nonlinear correction} spectra. Doing so we identify the optimal emulator configuration leading to a maximal emulation--only error of 0.27\% within the $0.83\Delta$ region of the parameter space. \gls{cod:EuclidEmulator}, constructed from 200 N-body simulations with this optimal configuration, almost perfectly reproduces the results of N-body simulated power spectra at the 0.3\% level within 50 ms. Due to possible numerical systematics in the N-body simulations themselves, the absolute accuracy of nonlinear power spectra generated with \gls{cod:EuclidEmulator} is bounded by $\pm 1\%$ up to $k\sim 6\hompc$ at $z=0$ while at \corrTwo{$z\sim 1$} this only holds up to $k\sim 1\hompc$ \citep{Schneider2015}. \corrOne{In order to reduce uncertainties due to dark matter physics that potentially contaminate studies of baryonic effects which are dominant at these scales,} it is hence vital in the future to further improve our confidence in the N-body simulations in the interval $1\hompc \leq k \leq 10\hompc$ for redshifts up to $z\sim3$ or higher. Once this can be achieved, thanks to the emulator strategy presented in this paper, the same accuracy will be reflected by the emulated power spectra. For Euclid these scales are important to assess the constraining power of the mission.

Our modeling approach will allow us to optimize the configuration of future emulators for further observables such as the bispectrum and the halo mass function, projecting their end-to-end accuracy, prior to running 
any simulations. We will also optimize emulation over a widened parameter space, adding neutrino mass, dark energy equation of state evolution, and primordial non-Gaussianity. These shall be included self-consistently 
within a future set of N-body simulations, with a mass resolution comparable to the Euclid Flagship simulation ($2.5 \times 10^9\; h^{-1} M_\odot$).

\corrOne{We have shown in \autoref{perf} that the run time of an emulation of a fully non-linear power spectrum with \gls{cod:EuclidEmulator} is highly dominated by the evaluation of the linear part with the Boltzmann solver. This motivates the need for a future, separate emulator of the linear power spectrum in order to speed up the entire process. Two separate emulators for the linear power spectrum and its non-linear correction are expected to perform better than one emulator for the non-linear power spectrum, as the separation approach allows a denser sampling of the parameter space in the construction of the experimental design leading to more accurate results.
}

\gls{cod:EuclidEmulator} can be downloaded from GitHub (\url{https://github.com/miknab/EuclidEmulator}). The repository contains the main C source code together with a python wrapper, CMake files, scripts and parameter files. Executing these scripts, the user can create a fully nonlinear power spectrum using \gls{cod:CLASS}\footnote{The \gls{cod:CLASS} code has to be installed separately. It can be downloaded from \url{http://class-code.net}.} and \gls{cod:EuclidEmulator}.

\section*{Glossary}
\label{glossary}
\glsfindwidesttoplevelname
\setglossarystyle{alttree}
\printglossary[type=main,title=Codes:]
\printglossary[type=acronym,title=Acronyms:]
\nopagebreak

\section*{Acknowledgements}
\label{ackns}
We express our gratitude to Julien Lesgourges whose feedback and assistance in the \gls{cod:CLASS} studies were highly appreciated. MK acknowledges support from the Swiss National Science Foundation (SNF) grant 200020\_149848. LL acknowledges support from the ERC starting grant ERCStg-717001. Simulations were performed on the zBox4+ cluster at the University of Zurich. The Euclid Consortium acknowledges the European Space Agency and the support of a number of agencies and institutes that have supported the development of Euclid. A detailed complete list is available on the Euclid web site (\texttt{http://www.euclid-ec.org}). In particular the Academy of Finland, the Agenzia Spaziale Italiana, the Belgian Science Policy, the Canadian Euclid Consortium, the Centre National d'Etudes Spatiales, the Deutsches Zentrum f\"ur Luft- and Raumfahrt, the Danish Space Research Institute, the Funda\c{c}\~{a}o para a Ci\^{e}nca e a Tecnologia, the Ministerio de Economia y Competitividad, the National Aeronautics and Space Administration, the Netherlandse Onderzoekschool Voor Astronomie, the Norvegian Space Center, the Romanian Space Agency, the State Secretariat for Education, Research and Innovation (SERI) at the Swiss Space Office (SSO), and the United Kingdom Space Agency.




\bibliographystyle{mnras}
\bibliography{Mendeley.bib} 



\appendix
\section{Principal components of the Experimental Design}
\label{app:PCs}
In contrast to \citet{Heitmann2010} we find that eleven principal components should be taken into account in order to bring the \gls{EOE} to a sub-percent level (see discussion in \autoref{Configuration}). In \autoref{fig:PCs} we plot the mean and the first eleven principal components (at $z=0$) of the \corrOne{nonlinear correction} spectra used for the construction of the \gls{cod:EuclidEmulator}. Notice that the emulation is performed using the logarithm of the nonlinear correction. This is why we report the mean and the corrections to the mean of $\ln(B)$. Recall further that the data presented is normalized.
\begin{figure*}
	\includegraphics[width=\textwidth]{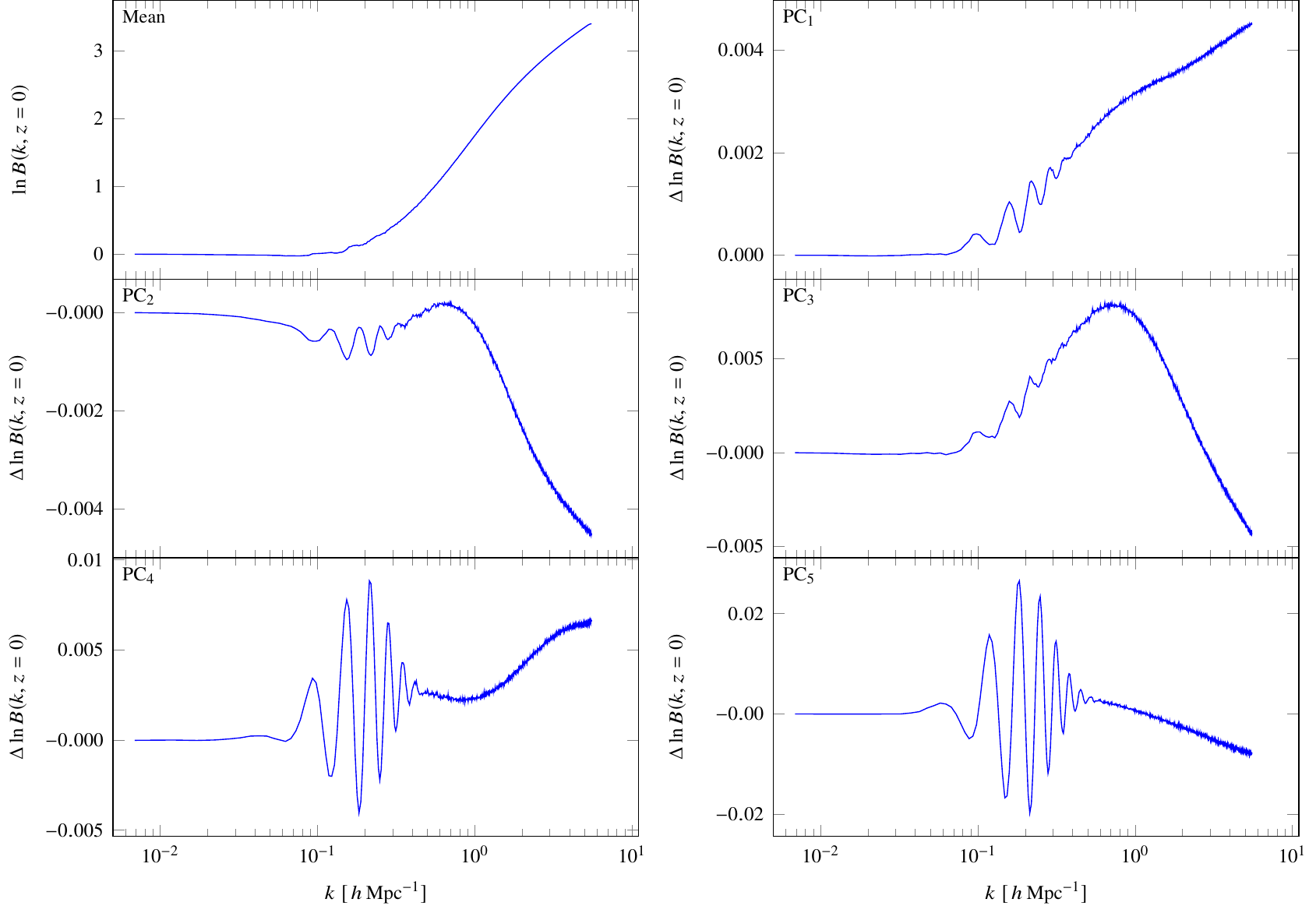}
	\caption{The mean and the first five principal components of the experimental design response data (logarithm of the \corrOne{nonlinear correction} spectra) used to construct the \gls{cod:EuclidEmulator}. Each of these curves is multiplied with its PCA coordinate $\lambda$ which is the actual output of the S\gls{PCE} emulator. The sum of these then produces the final \corrOne{nonlinear correction} spectrum.}
    \label{fig:PCs}
\end{figure*}
\begin{figure*}
	\includegraphics[width=\textwidth]{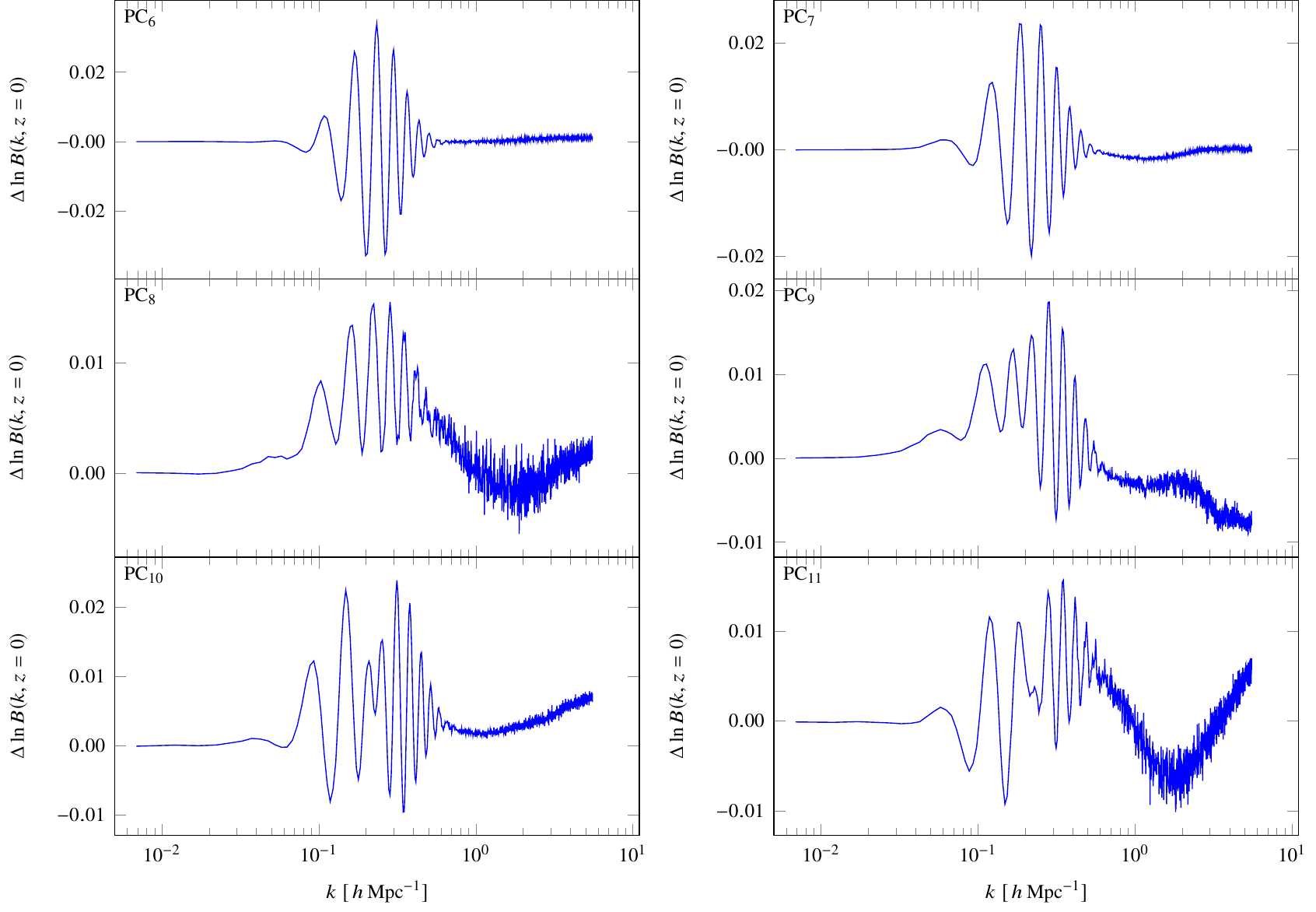}
	\contcaption{Principal components ${\rm PC}_6$ to ${\rm PC}_{11}$ of the experimental design response data (logarithm of the \corrOne{nonlinear correction} spectra) used to construct the \gls{cod:EuclidEmulator}. ${\rm PC}_{11}$ is the highest order principal component used in the \gls{cod:EuclidEmulator}}
\end{figure*}

For each principal component, a Sobol' index analysis (see \autoref{Sensitivity}) can be performed. The results for the first and the fifth \gls{PC} are shown in \autoref{SobolPlot}, which tells us, that ${\rm PC}_{1}$ is sensitive to all cosmological parameters but $\omega_{\rm b}$, while ${\rm PC}_{5}$ mostly depends on $\omega_{\rm m}$ and $h$.

\section{Surrogate modeling via SPCE}
\label{App:UQ}
\subsection{Introduction to surrogate modeling}
The ability to make predictions and to properly propagate the input uncertainties to the output response vector in cases of complex systems is of prime interest in numerous situations. Yet, it is infeasible to perform expensive large-scale experiments or simulations for many input parameter sets to study the system's behavior in detail. In such cases a surrogate model (or ``emulator'') can be computed that establishes a (model) relation between input and output. This means that the surrogate is not the ``true'' relation but, depending on the computational resources available, it is able to capture the main features of this ``true'' relation up to a required accuracy. The uncertainties can be kept under control using well-known techniques from statistical uncertainty quantification.

Mathematically we can formulate the problem of emulating a black-box model as follows:  consider a set of input parameters as a random vector ${\mathbf X}\in\mathbb{R}^d$
\begin{equation*}
{\vX}:=\{X_i\vert i\in 1,...,d; d\in \mathbb{N}\}\,,
\end{equation*}
where $d$ parametrizes the dimensionality of the parameter space under consideration. The probability distributions of each of the independent components of $\vX$ are given by: $X_i \sim F_{X_i}(x_i)$. These parameters are mapped by a black-box relation $\cm$ (the computational model) to a quantity of interest $Y$: 
\begin{equation*}
Y:=\cm(\vX)
\end{equation*}
with $Y\in\mathbb{R}$. Due to the uncertainty in the input vector $\mathbf{X}$, $Y$ is also a random variable. 

The goal is to find a surrogate model $\mathcal{S}$ relating $\mathbf{X}$ to $Y$ based on a
small set of model evaluations known as the ``experimental design'' (ED) 
${\cx = \left\{ \vx^{(1)},\cdots,\vx^{(n_{\rm ED})} \right\}}$ 
and the corresponding model responses 
$\cy = \left\{y^{(1)}=\cm\left(\vx^{(1)}\right),\cdots,y^{(n_{\rm ED})}=\cm\left(\vx^{(n_{\rm ED})}\right)\right\}$. 

In realistic scenarios, the computational budget to create an \gls{ED} is limited and this limitation puts a constraint on the amount of information one can use to construct the surrogate model. Further we add the requirement that the surrogate model $\mathcal{S}$ must be as accurate as possible throughout the entire parameter space spanned by $\vX$.\\

In this section we will focus on a specific type of surrogate model, polynomial chaos expansions \citep{Wiener1938TheChaos}, in contrast to \citet{Heitmann2010} and \citet{Lawrence2017} where they use Kriging for both the {\em FrankenEmu} and the \gls{cod:CosmicEmu}. \corrOne{We have chosen the polynomial chaos expansion approach to construct EuclidEmulator bacuase globally the errors are expected to be smaller than in the case of Kriging. Further, a \gls{PCE} approach allows to relax assumptions about the noise properties of the input model: Kriging can indeed deal with noise, but a very severe assumption on the noise distribution is to be made; the noise is assumed to be Gaussian. In a linear regression setting such as \gls{PCE}, the only assumption is that the noise is unbiased (see e.g. \citealt{Vapnik1998Statistical1998}).}

\subsection{Theory of polynomial chaos expansions}
\label{sec:Theory}
The concept of polynomial chaos expansion (\gls{PCE}) based on \gls{LARS} shall quickly be reviewed here. For a deeper discussion of this topic we refer to \citet{Blatman2009AdaptiveAnalysis,Blatman2009SparseAlgorithm, Blatman2010AnAnalysis, Blatman2011AdaptiveRegression} and references therein.

Let us take $\mathbf{X}=\{X_1, \dots, X_d\}^\top\in \mathbb{R}^d$ to be a random input vector with joint probability density function (pdf) $f_\mathbf{X}(\vx)$ and a finite variance model $\mathcal{M}$ mapping $\mathbf{X}$ to the response $Y$ via $Y:=\mathcal{M}(\mathbf{X})$, i.e.
\begin{equation}
\mathbb{E}[Y^2]=\int_\mathcal{D_\mathbf{X}}\mathcal{M}^2(\vx)f_\mathbf{x}(\vx){\rm d}\vx < \infty\,,
\end{equation}
where $\mathcal{D}_\mathbf{x}$ is the domain of the random input vector. Then $\cm(\vX)$ is an element of the stochastic Hilbert space $\mathcal{H}$ of finite variance functions endowed with the inner product
\begin{equation}
\label{eqn:InnerProd}
\left\langle g,h\right\rangle := \mathbb{E}\left[g(\vX)\cdot h(\vX)\right] = \int_\mathcal{D_\mathbf{X}} g(\vX) h(\vX) f_{\vX}(\vx){\rm d}\vx\,.
\end{equation}
Then the following spectral representation, known as polynomial chaos expansion, holds:
\begin{equation}
\label{PCInf}
Y = \cm(\vX) = \sum\limits_{\mathbf{\alpha}\in \mathbb{N}^d} \etaalpha\Psialpha (\vX)\,,
\end{equation}
where $\valpha = \{\alpha_1,\cdots,\alpha_d\}$ is a multi-index, $\Psialpha$ is an element of a multivariate orthonormal polynomial basis of $\mathcal{H}$ and $\etaalpha$ is the corresponding coefficient (coordinate). The multi-variate polynomials $\Psialpha$ are constructed by tensor products of univariate orthonormal polynomials w.r.t. the input random variables:
\begin{equation}
\label{eqn:PsiAlpha}
\Psialpha(\vX) = \prod\limits_{i = 1}^d \phi^{(i)}_{\alpha_i}(x_i)\,,
\end{equation}
where $\phi^{(i)}_{\alpha_i}(x_i)$ is a polynomial of degree $\alpha_i$ in $x_i$ orthonormal w.r.t. the pdf of $f_{X_i}(x_i)$. In other words:
\begin{equation}
\label{eqn:phiortho}
\left\langle \phi_\alpha, \phi_\beta \right\rangle = \delta_{\alpha\beta}\,.
\end{equation}
From \autoref{eqn:PsiAlpha}, it follows that the total degree of the basis element $\Psialpha(\vX)$ is $ p = ||\alpha||_1 = \sum\limits_{i = 1}^d \alpha_i$, while from \autoref{eqn:phiortho} it follows that for an input random vector with independent components $\vX$:
\begin{equation}
\label{eqn:psiortho}
\left\langle \Psialpha, \Psi_\mathbf{\beta} \right\rangle = \delta_{\valpha\mathbf{\beta}}\,.
\end{equation}

In the present case, \gls{cod:EuclidEmulator} is built to be consistently accurate across predefined parameter intervals. Therefore, their input distributions are considered uniform between the given bounds. Prior to the expansion, each parameter is linearly rescaled to the interval $[-1, 1]$, so that the polynomials $\phi^{(i)}_{\alpha_i}$ used in the expansions belong to the Legendre family \citep{Ghanem2003StochasticApproach,Xiu2006TheEquations}.

\subsubsection{Truncation of the polynomial basis}
For practical computational purposes, the expansion in  \autoref{PCInf} needs to be truncated to a finite number of terms:
\begin{equation}
Y = \cm(\vX) \approx \sum\limits_{\mathbf{\alpha}\in \ca} \etaalpha\Psialpha (\vX)\,,
\end{equation}
where $\ca$ is a truncation with cardinality ${P := \text{card}(\ca) < \infty}$.

Several strategies are available to define a suitable truncation set $\ca$ in the literature. The scheme applied in the construction of the \gls{cod:EuclidEmulator} is a combination of the so called \textit{maximum interaction} and \textit{hyperbolic truncation} introduced in \citet{Blatman2011AdaptiveRegression,Marelli2017UQLabExpansion}.

The \textit{standard truncation scheme} is given by retaining only basis functions up to a specific total degree $p$ such that 
\begin{equation}
\mathcal{A}^{d,p} = \{\alpha\in\mathbb{N}^d, \vert\vert\alpha\vert\vert_1\leq p\}\,.
\end{equation}
The cardinality of such a set is:
\begin{equation}
\text{card}(\mathcal{A}^{d,p}) = \begin{pmatrix} d+p\\p\end{pmatrix}
\end{equation}
which is a polynomially increasing quantity. A significant reduction of the number of basis elements is to impose bounds on the maximum number of non-zero elements in $\alpha_i$ to a desired $r \leq d$: 
\begin{equation}
\mathcal{A}^{d,p,r} = \{\alpha\in\mathcal{A}^{d,p}, \vert\vert\alpha\vert\vert_0<r\}\,,
\end{equation}
where
\begin{equation}
\vert\vert\alpha\vert\vert_0:=\sum_{i=1}^d1_{\alpha_i>0}
\end{equation}
is the rank of the multi-index. 
The effect of this is that in each multivariate polynomial chaos basis function only $r$ or less univariate factors are not constant and hence $r$ or less input parameters interact with each other (a ``maximum interaction'' is defined).

We reduce the number of terms taken into account once more by applying hyperbolic truncation. This is closely related to the standard truncation scheme with the difference that instead of the 1-norm a more general q-norm is used with $q\in[0,1]$:
\begin{equation}
\mathcal{A}^{d,p,q} = \{\alpha\in\mathcal{A}^{d,p}, \vert\vert\alpha\vert\vert_q\leq p\}\,,
\end{equation}
where 
\begin{equation}
\vert\vert\alpha\vert\vert_q:=\left(\sum_{i=1}^d\alpha_i^q\right)^{1/q}\,.
\end{equation}

Hence, for a hyperbolic and maximum interaction limited truncation we get:
\begin{equation}
\mathcal{A}^{d,p,q,r} = \{\alpha\in\mathcal{A}^{d,p}, \vert\vert\alpha\vert\vert_q\leq p\;\text{and}\;\vert\vert\alpha\vert\vert_0<r\}\,.
\end{equation}
Notice that only $d$ is specified as it is the dimension of the input random vector $\mathbf{X}$. The maximal polynomial order $p$ can be found automatically following the iterative approach described in detail in \citet{Blatman2011AdaptiveRegression}. Finding the optimal values for $q$ and $r$, on the other hand, requires a dedicated parametric study, discussed in \autoref{ErrorPrediction}.

\subsubsection{Calculating the \gls{PCE} coefficients with sparse regression}
\label{app:LARS}
Once the polynomial basis has been constructed, the expansion coefficients $\eta_\valpha$ need to be calculated.
Given the high computational costs of the computational model, \gls{cod:EuclidEmulator} employs the sparse-regression approach in \citet{Blatman2011AdaptiveRegression}, based on the well known \textit{least-angle regression} technique first introduced in \citet{Efron2004LEASTREGRESSION}. 
This approach has been widely demonstrated to be highly efficient even in the presence of high dimensional or highly nonlinear models, as it favors highly sparse models so as to avoid over-fitting in the presence of small experimental designs.

\corrOne{
\subsubsection{A note on the extrapolation properties of PCE}
As mentioned already above, \gls{PCE} is not an interpolant but a regression technique. This means that a \gls{PCE}-based surrogate model is able to accurately estimate the response of the input model not just near the positions of the experimental design points but also further away from them (on a global scale). However, due to the fact that the cosmological parameters have to be mapped to the interval $[-1,1]$ in order to be evaluated by the Legendre polynomials (see explanations in \autoref{subsec:surmodel}), the regression only works within the predefined parameter bounds. If one wants to predict the response for a cosmology outside the input bounds, a new emulator has to be trained. This will result in different basis functions and coefficients.
}

\section{Simulations and Convergence Tests}
\label{App:ConvTests}
Since there is no analytical way to compute a ``true'' nonlinear power spectrum, a convergence test for the power spectrum is necessary. In \autoref{SimTable} we list all simulations we have used in this work together with their specifications and the required runtime $\Delta T$ in node hours. We assign a unique label to each simulation that we use for reference in the text below. We define $L$ to be the length of a simulation box edge in units of $\mpcoh$, $N$ denotes the number of particles per dimension used in a simulation to create the initial conditions and $R_{\rm grid}=N_{\rm ma}/N$ is the ratio between the number of cells $N_{\rm ma}$ used for the mass assignment and the number of particles $N$.

\begin{table*}
\centering%
\caption{
List of simulations used in this paper. For each simulation, its unique ID as well as its specifications are listed. The specifications consist of the box size ($L$), the number of particles per side length ($N$), whether it is a paired-and-fixed run (PF yes/no) and the runtime in node hours. The simulation 000 is a \gls{cod:EuclidEmulator}-like simulation but run based on the Euclid reference cosmology, which is not part of the \gls{ED}. Simulations 001 to 100 are the runs that form the actual \gls{cod:EuclidEmulator}-ED while 101 to 136 were used for the end-to-end test reported in \autoref{E2E}. \gls{EFHR} and \gls{LV} denote the two reference simulations for the convergence tests against which the simulations CT${}_1$ to CT${}_{20}$ were compared. For all CT-runs with $N=1024$ the power spectra were measured with three different mass assignment grids (see \autoref{fig:LvsR}), indicated by the three labels a, b and c (this did not require separate simulations). Note that the \gls{EFHR} simulation was run with GPUs. The total run time for all simulations sums up to over 380\,000 node hours.
}
\begin{tabular}{lcccccc}
Simulation identifier&$L$ $[\mpcoh]$&$N$&PF& number of runs & total runtime [node hours] \\
\hline
\hline
000&1250&2048&yes&2&1\,904\\
001-100&1250&2048&yes&200&190\,200\\
101-136&1250&2048&yes&72&68\,472\\
\hline
Euclid Flagship High Resolution (EFHR)&1920&8000&no&1&93\,600${}^\star$\\
Large Volume (LV)&4000&4096&yes&2&14\,696\\
\hline
Convergence Test 1 (CT${}_1$)&256&262&yes&2&23\\
CT${}_2$&512&524&yes&2&50\\
CT${}_3$&640&655&yes&2&80\\
CT${}_4$&960&983&yes&2&236\\
CT${}_5$&1250&1280&yes&2&508\\
CT${}_{6\rm abc}$ (a: $R_{\rm grid}=1$, b: $R_{\rm grid}=2$, c: $R_{\rm grid}=7$)&480&1024&yes&2&378\\
CT${}_7$&480&1536&yes&2&1\,402\\
CT${}_{8\rm abc}$&640&1024&yes&2&308\\
CT${}_{9}$&640&1536&yes&2&1\,204\\
CT${}_{10\rm abc}$&960&1024&yes&2&240\\
CT${}_{11}$&960&1536&yes&2&896\\
CT${}_{12}$&960&2048&yes&2&2\,274\\
CT${}_{13\rm abc}$&1440&1024&yes&2&209\\
CT${}_{14}$&1440&1536&yes&2&752\\
CT${}_{15}$&1440&1920&yes&2&1\,953\\
CT${}_{16\rm abc}$&1920&1024&yes&2&184\\
CT${}_{17}$&1920&1536&yes&2&633\\
CT${}_{18}$&1920&1920&yes&2&1529\\
\hline
&&&&&\\
${}^\star$ with GPUs&&&&&\\
\label{SimTable}
\end{tabular}
\end{table*}

\subsection{Simulation parameters}
The goal in this work is to find the minimal volume, number of particles and mass assignment grid size that allows us to achieve the required 1\% accuracy over the $k$ range of interest. A number of further parameters like softening, time-stepping have already been assessed in \citet{Schneider2015}. They report that varying the time-stepping and softening parameters has a sub-percent effect over all $k$-scales of interest. We use the \gls{cod:PKDGRAV3}-default values which have been shown to be reasonable choices: the softening is given by $\epsilon = 0.02\;l_{\rm mean}$, with $l_{\rm mean}$ being the mean inter-particle distance. The time-stepping parameter $\eta=0.2$ controls each individual particle's time step via, $\Delta T = \eta\sqrt{\epsilon/a}$ with $a$ being the gravitational acceleration of the particle.

In what follows we focus on the box volume and mass resolution (i.e. particle number) as well as the size of the mass assignment grid. We perform the convergence test in three steps: first, we determine the minimal simulation box volume by comparing to a paired-and-fixed simulation in large volume (\gls{LV}) of $(4000\;\mpcoh)^3$ with $4096^3$ particles. Secondly, we find the minimal mass resolution by converging toward an extreme high resolution run (Euclid Flagship High Resolution, \gls{EFHR}) with $8000^3$ particles in a $(1920\;\mpcoh)^3$-box \corrOne{($N_{\rm ma}^{\rm EFHR}=8000$)} and thirdly, the minimal size of the mass assignment grid is assessed.

\begin{figure*}
	\includegraphics[width=\textwidth]{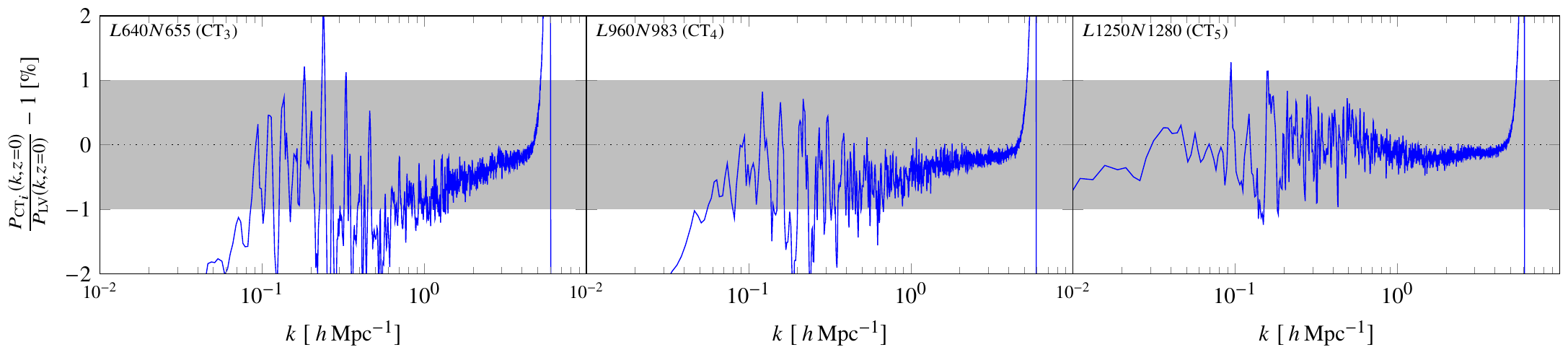}
	\caption{
     Comparison of three different runs with different box sizes but equal mass resolution $\linv=1.024\hompc$ against a large volume (\gls{LV}) reference simulation with a box of $L=4000\mpcoh$ side length and $N=4096$ particles per dimension. Only simulation volumes of at least ${L^3=(1250\;\mpcoh)^3}$ (right panel) allow a power spectrum measurement that agrees with the one in a large volume at the 1\% level. Smaller simulation volumes lead to a power deficit on large scales (left and middle panel) and increasingly larger deviations from the reference on mildly nonlinear scales.
    }
    \label{fig:VolConvTest}
\end{figure*}

For the minimal volume measurement, we compare five paired-and-fixed runs
\begin{itemize}
\item ${L=256\;\mpcoh}$ and ${N=262}$ (CT${}_1$),
\item ${L=512\;\mpcoh}$ and ${N=524}$ (CT${}_2$),
\item ${L=640\;\mpcoh}$ and ${N=655}$ (CT${}_3$),
\item ${L=960\;\mpcoh}$ and ${N=983}$ (CT${}_4$) and 
\item ${L=1250\;\mpcoh}$ and ${N=1280}$ (CT${}_5$)
\end{itemize}
against the \gls{LV}-simulation (notice that the mass resolution is the same for all these simulations). According to \autoref{fig:VolConvTest}, we find that a minimal simulation box volume of ${L^3=(1250\;\mpcoh)^3}$ is necessary for the power spectrum to converge to the \gls{LV}-power spectrum to within 1 percent at large scales. This result \corrOne{is consistent with the very recent paper \cite{Klypin2018DarkMatrices} and} updates the conclusion drawn in \citet{Schneider2015} where they claim a lower bound for cosmological simulation box sizes of only $500\mpcoh$. A potential reason for this underestimation is that their reference simulation volume is only $1024\mpcoh$ and hence most likely too small.

In the second step, we determine the minimal mass resolution, which we measure in terms of $\linv:=N/L$ corresponding to the inverse of the mean inter-particle separation. We define a discrete parameter space by
\begin{align*}
L&\in\{480,640,960,1440,1920\}\mpcoh\,,\\
N&\in\{1024,1536,2048\}\;,\\
R_{\rm grid}&\in\{1,2,4\}
\end{align*}
and run N-body simulations for three selected planes in this space:
\begin{itemize}
\item the $L$-$N$-plane with $R_{\rm grid}=1$,
\item the $L$-$R_{\rm grid}$-plane with $N=1024$ and
\item the $N$-$R_{\rm grid}$-plane with $L=960\mpcoh$\;.
\end{itemize}
We perform this convergence test against the \gls{EFHR}-simulation ($L=1920\;\mpcoh$ and $N=8000$, $\linv_{\rm EFHR}=4.167$). 
\begin{figure*}
	\includegraphics[width=\textwidth]{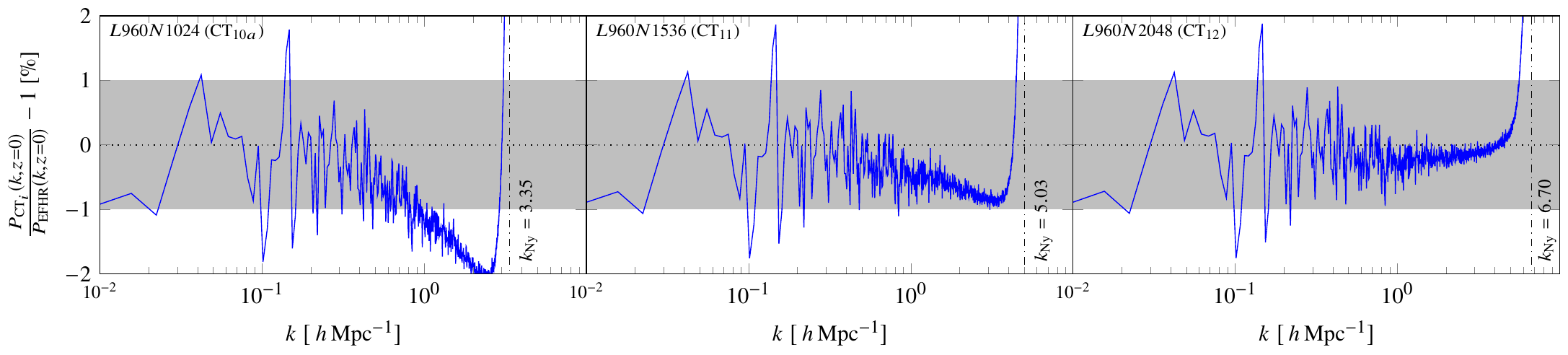}
	\caption{Effect of measuring the power spectrum with different \corrOne{mass resolutions} and with different number of particles ($R_{\rm grid}=1$). \corrOne{Shown is a subset of the $L$-$N$-plane with fixed $L$: all panels belong to the same box size $L=960 \mpcoh$.} From left to right, $N=1024$, 1536 and 2048 particles per dimension were used. We thus span an inverse mean inter-particle distance range of ${1.07\hompc\leq \linv \leq 2.13\hompc}$. The turn-up of the error curve at high $k$-modes indicates the location of the Nyquist frequency (dash-dotted line) of the mass assignment grid.}
    \label{fig:LvsN}
\end{figure*}
Increasing the number of particles $N$ used in a simulation of a given volume, we reduce aliasing that comes from discreteness. This is expected because in the limit $N\to\infty$ we approximate the real fluid case we are actually interested in. We observe that a resolution parameter of $\linv=2048/960=2.13 \hompc$ (\autoref{fig:LvsN}, right panel) yields almost perfect results on highly nonlinear scales (up to the point where the Nyquist effect from the mass assignment grid kicks in) while a resolution of only $\linv=1536/960=1.6\hompc$ (\autoref{fig:LvsN}, middle panel) leads to an aliasing artifact that only just stays within the 1\% region. From this we conclude that in order to meet the 1\% accuracy level required by Euclid over all scales of interest we need a resolution parameter $\linv\geq1.6\hompc$.

\begin{figure*}
	\includegraphics[width=\textwidth]{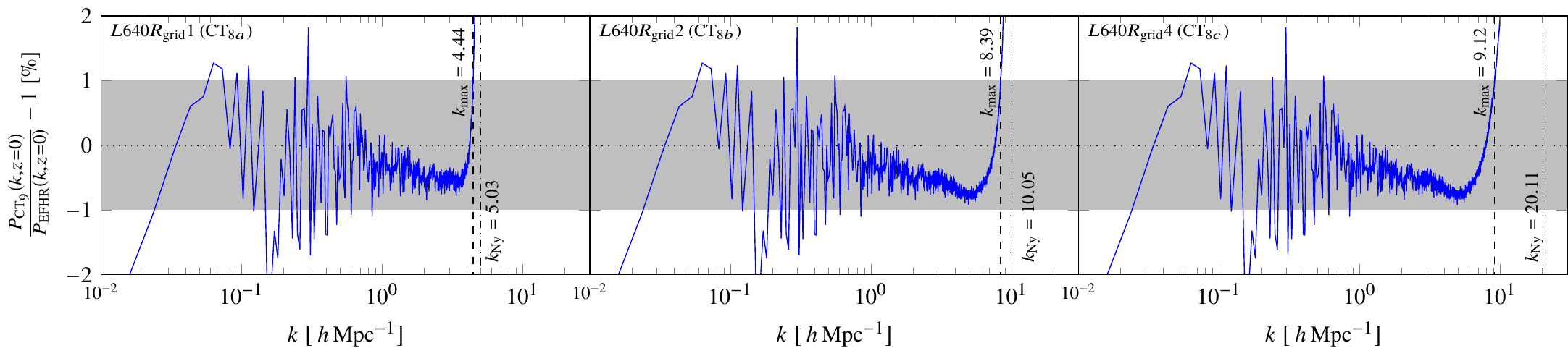}
	\caption{Effect of measuring the power spectrum using different grid sizes, i.e. in the $L$-$R_{\rm grid}$-plane. From left to right, the columns refer the number of sampling grid points being equal to once, twice and four times the number of points in the particle grid per dimension, respectively. The number of particles is $N=1024$ in all subplots. The dashed vertical lines indicate $k_{\rm max}$ (the frequency at which the relative error becomes larger then $1\%$) while the dash-dotted lines again show the Nyquist frequency of the mass assignment grid.}
    \label{fig:LvsR}
\end{figure*}

We are then left with assessing how small the mass assignment cells have to be in order to reach the desired precision. We performed another set of simulations for the five different box sizes where the mass assignment cells are either $1$, $1/8$ or $1/64$ times the volume of the particle grid cells, respectively (cf. \autoref{fig:LvsR}). The Nyquist frequency $f_{\rm Ny}$ of the mass assignment grid itself is linearly proportional to $N_{\rm ma}$ (and thus also to $R_{\rm grid}$ for a given $N$) according to the Shannon-Nyquist theorem \citep{Nyquist1928CertainTheory}
\begin{equation}
\label{eq:NyquistTheorem}
f_{\rm Ny}=\frac{1}{2}N_{\rm ma}\frac{2\pi}{L}=\frac{1}{2}R_{\rm grid}N\frac{2\pi}{L}\,,
\end{equation} 
where $2\pi/L$ corresponds to the canonical inter-particle scale $\Delta k$ used by the fast Fourier transform (FFT) such that $N_{\rm ma}\Delta k$ equals the maximal Fourier mode for the power spectrum measurement. It is not a priori clear, though, by what factor an increasing value of $R_{\rm grid}$ increases the $k$-interval within which the error curve remains bounded by 1\%. 
Let us define $k_{\rm max}$ to be the maximal $k$-value such that
\begin{equation}
100\left(\frac{P_{{\rm CT}_i}(k,z=0)}{P_{\rm EFHR}(k,z=0)}-1\right)>1,\quad\forall k>k_{\rm max}\,.
\end{equation}
We find that the proportionality given in \autoref{eq:NyquistTheorem} translates rather well to a proportionality between $k_{\rm max}$ and $R_{\rm grid}$ for low resolution parameters and small values of $R_{\rm grid}$ while for high $\linv$ and larger values of $R_{\rm grid}$ it breaks down as the slope of the Nyquist turn-up becomes more and more shallow for larger values of $R_{\rm grid}$. For instance, while we can essentially double $k_{\rm max}$ in a $L$640$N$1024 simulation by going from $k_{\rm max}(R_{\rm grid}=1)=4.4\hompc$ (\autoref{fig:LvsR}, left panel) to $k_{\rm max}(R_{\rm grid}=2)=8.39\hompc$ (\autoref{fig:LvsR}, center panel), we cannot do so again by increasing $R_{\rm grid}$ from 2 to 4 (\autoref{fig:LvsR}, right panel). In this figure, data from a 
$L$640$N$1024 simulation is shown which fulfills the resolution condition of $\linv=1.6\hompc$ found above. One can see that the error stays within 1\% up to $k_{\rm max}(R_{\rm grid}=1)=4.44\hompc$, $k_{\rm max}(R_{\rm grid}=2)=8.39\hompc$ and $k_{\rm max}(R_{\rm grid}=4)=9.12\hompc$. Taking into account that increasing $R_{\rm grid}$ leads to a non-negligible increase in required computational resources, we conclude that $R_{\rm grid}=2$ is a reasonable choice. 

We thus end up with the following lower bounds:
\begin{equation}
\begin{split}
\linv&\geq 1.6\hompc\,,\\
R_{\rm grid}&\geq 2\,.&
\end{split}
\end{equation}
Summarizing the results of our convegence tests, we decide to use the following specifications for the construction of the \gls{cod:EuclidEmulator} experimental design:
\begin{equation}
\label{Requirements:Equ:SimPars}
\begin{split}
L&=1250\mpcoh\,,\\
N&=2048\,,\\
R_{\rm grid}&= 2\,.
\end{split}
\end{equation}
Notice that this choice obeys the constraint put on $\linv$ as here $\linv=1.638\hompc$. This choice of simulation volume and particle number corresponds to a mass resolution of roughly $2\times10^{10}\;h^{-1} M_\odot$ per particle.

\subsection{Redshift dependence of \texorpdfstring{$k_{\rm max}$}{kmax}}
The chosen configuration for the \gls{ED} simulations suggest that at $z=0$ the simulated nonlinear power spectrum can be trusted up to $k_{\rm max} \approx 8\hompc$. Of course, in order to be able to produce a reliable \corrOne{nonlinear correction} prediction using the \gls{cod:EuclidEmulator} at a certain redshift, it is of utmost importance to know how $k_{\rm max}$ changes with redshift. 
We have found that the initial power spectrum at $z=200$ measured by \gls{cod:PKDGRAV3} has converged to \corrOne{linear theory (as computed by CLASS)} up to $k_{\rm max}=5.48\hompc$. The convergence test in the previous section suggests that the nonlinear power spectrum at lower $z$ converge up to larger $k_{\rm max}$ (for $z=0$ we find $k_{\rm max}\approx 8\hompc$). The division by the initial power spectrum in the computation of the \corrOne{nonlinear correction} renders the latter to be converged up to $k_{\rm max}=5.48\hompc$ for all redshifts up to $z = 5$. Based on this the allowed redshift and $k$ range for emulation with the \gls{cod:EuclidEmulator} is set to $0\leq z \leq 5$ and ${0.01\hompc\leq k\leq 5\hompc}$. For $z>5$ we found a non-trivial dependence of $k_{\rm max}$ on the redshift. However, it is not clear yet to what extent this functional relation is influenced by numerical artifacts (like e.g. aliasing or transients) and to what extent it is physical.

\corrOne{
\section{Pairing-and-fixing vs. Gaussian initial condition-based simulations}
\label{App:PF}
In this appendix we address potential issues of pairing-and-fixing (introduced in \citealt{Pontzen2016InvertedVoids}) and compare emulated nonlinear corrections to nonlinear corrections coming from a traditional, Gaussian initial condition sample. In \citet{Angulo2016} it is explained that fixing the power spectrum amplitude in the initial conditions allows to approximate the ensemble mean of a set of power spectra with Gaussian initial conditions with no variance at the cost of introducing some non-Gaussianity into the initial conditions. As is shown in the lower panel of Fig. 2 in \citet{Angulo2016}, a deviation of the PF mean from the ensemble mean of power spectra can be observed at high $k$ but it stays inside the 0.1\% region up to $k\gtrsim 1 \hompc$.
}

\begin{figure}
	\includegraphics[width=\columnwidth]{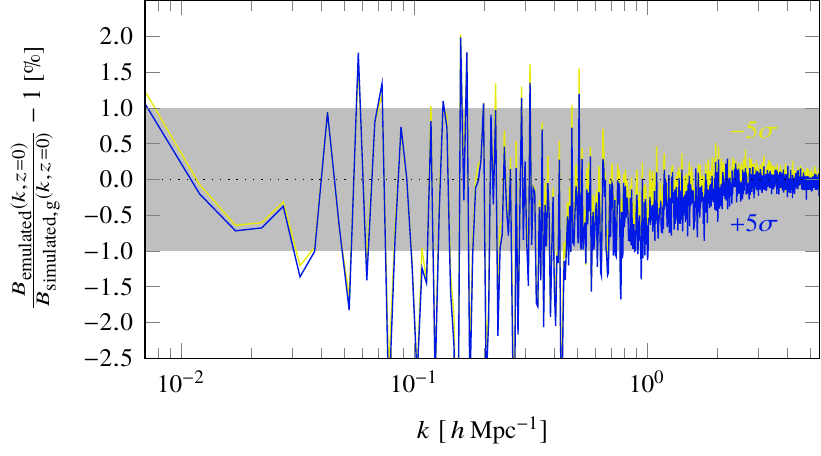}
	\caption{\corrOne{A similar comparison as in \autoref{fig:hcompPlot}. Shown is the relative difference between the emulated \corrOne{nonlinear corrections} and those computed from Gaussian random field initial condition-based simulations (for the two extreme cases ``$+5\sigma$'' and ``$-5\sigma$''). The high-frequency errors are due to computational cosmic variance which (in our case) mainly has an effect on intermediate scales.}}
    \label{fig:EEvsGaussian}
\end{figure}

\corrOne{\autoref{fig:EEvsGaussian} is a plot similar to \autoref{fig:hcompPlot} with the difference that here we compare to the nonlinear correction of a single run with Gaussian random field initial conditions (for the ``$+5\sigma$'' and the ``$-5\sigma$'' case; we use the same notation as was used in \autoref{fig:hcompPlot}). We find that on large scales the computational cosmic variance does not play a big role. This is due to the fact that in order to compute the nonlinear correction, one divides the power spectrum at a given redshift $z$ by the initial condition of the simulation and this cancels out most of the variance. However, the biggest deviations are observed on mildly nonlinear scales. On these intermediate scales, the variation is initially small but is amplified nonlinearly during the evolution. Hence, division by initial condition is not enough to efficiently cancel it. Pairing and fixing does decrease the cosmic variance on these scales to some degree. \citet{Villaescusa-Navarro2018StatisticalFields} have studied the $k$-dependence on how much paired-and-fixed simulations feature less computational cosmic variance compared to usual Gaussian random field simulations. They find that towards smaller scales, the improvement brought by pairing-and-fixing degrades to the point where the pairing-and-fixing approach performs equally well as the classical Gaussian random field approach. In the context of our work, however, this is not a problem as on smaller scales cosmic variance does not play an important role in the first place.}

\bsp	
\label{lastpage}
\end{document}